\documentclass[useAMS,usenatbib]{mn2e}
\usepackage{rotating}
\usepackage{subfigure}
\RequirePackage{setspace}
\usepackage{graphicx}
\usepackage{fancyhdr}
\usepackage{natbib}
\usepackage{amsmath}
\usepackage{amssymb}
\usepackage{aas_macros}
\usepackage{epstopdf}
\usepackage{lscape}
\usepackage{rotating}
\usepackage{lineno} 
\usepackage{float} 

\title[GAMA: passive low-mass galaxies]{Galaxy And Mass Assembly (GAMA): growing up in a bad neighbourhood - how do low-mass galaxies become passive?}
\author[L. J. M. Davies et. al.]{L. J. M. Davies$^{1}$\thanks{E-mail:
 luke.j.davies@uwa.edu.au}, A. S. G. Robotham$^{1}$, S. P. Driver$^{1,2}$, M. Alpaslan$^{3}$, I. K. Baldry$^{4}$, \newauthor J. Bland-Hawthorn$^{5}$,  S. Brough$^{6}$, M. J. I. Brown$^{7}$ , M. E. Cluver$^{8}$, B. W. Holwerda$^{9}$,  \newauthor A. M. Hopkins$^{6}$,  M. A. Lara-L\'opez$^{10}$, S. Mahajan$^{11}$, A. J. Moffett$^{1}$,  M. S. Owers$^{6,12}$,  \newauthor S. Phillipps$^{13}$\\
 \\
$^{1}$ ICRAR, The University of Western Australia, 35 Stirling Highway, Crawley, WA 6009, Australia\\ 
$^{2}$ SUPA, School of Physics and Astronomy, University of St Andrews, North Haugh, St Andrews, Fife, KY16 9SS, UK\\
$^{3}$ NASA Ames Research Center, N232, Moffett Field, Mountain View, CA 94035, USA \\
$^{4}$ Astrophysics Research Institute, Liverpool John Moores University, IC2, Liverpool Science Park, 146 Brownlow Hill, Liverpool, L35RF, UK\\
$^{5}$ Sydney Institute for Astronomy, School of Physics A28, University of Sydney, NSW 2006, Australia\\
$^{6}$ Australian Astronomical Observatory, PO Box 915, North Ryde, NSW 1670, Australia\\
$^{7}$ School of Physics and Astronomy, Monash University, Clayton, Victoria 3800, Australia \\ 
$^{8}$ University of the Western Cape, Robert Sobukwe Road, Bellville, 7535, South Africa \\
$^{9}$ University of Leiden, Sterrenwacht Leiden, Niels Bohrweg 2, NL-2333 CA Leiden, The Netherlands\\
$^{10}$ Instituto de Astronom\'ia, Universidad Nacional Aut\'onoma de M\'exico, A.P. 70-264, 04510 M\'exico, D.F., M\'exico\\
$^{11}$ Indian Institute of Science Education and Research Mohali, Knowledge City, Sector 81, Manauli 140306, Punjab, India\\
$^{12}$ Department of Physics and Astronomy, Macquarie University, NSW 2109, Australia\\
$^{13}$ School of Physics, University of Bristol, Tyndall Avenue, Bristol BS81TL, UK\\
}
\begin{document}

\date{Draft: Nov 2015}

\pagerange{\pageref{firstpage}--\pageref{lastpage}} \pubyear{2015}

\maketitle

\begin{abstract}

Both theoretical predictions and observations of the very nearby Universe suggest that low-mass galaxies (log$_{10}$[M$_{*}$/M$_{\odot}$]$<9.5$) are likely to remain star-forming unless they are affected by their local environment. To test this premise, we compare and contrast the local environment of both passive and star-forming galaxies as a function of stellar mass, using the Galaxy and Mass Assembly survey. We find that passive fractions are higher in both interacting pair and group galaxies than the field at all stellar masses, and that this effect is most apparent in the lowest mass galaxies. We also find that essentially all passive log$_{10}$[M$_{*}$/M$_{\odot}$]$<8.5$ galaxies are found in pair/group environments, suggesting that local interactions with a more massive neighbour cause them to cease forming new stars. We find that the effects of immediate environment (local galaxy-galaxy interactions) in forming passive systems increases with decreasing stellar mass, and highlight that this is potentially due to increasing interaction timescales giving sufficient time for the galaxy to become passive via starvation. We then present a simplistic model to test this premise, and show that given our speculative assumptions, it is consistent with our observed results.

\end{abstract}

\begin{keywords}
galaxies: interactions - galaxies: evolution
\end{keywords}

\section{Introduction}
\label{sec:Intro}

A ubiquitous feature of galaxy evolution is the transformation of galaxies from blue actively star-forming systems into red passive systems \citep[$e.g.$][]{Tinsley68}. There has been significant debate about the factors affecting such transitions \citep[$e.g.$][]{Baldry06, Kimm09,Peng10,Wijesinghe12}, the likely timescales over which the transformation occurs \citep[$e.g.$][]{Balogh04,Martin07,Wetzel13,Wheeler14,Fillingham15} and the morphological changes associated with the quenching of star formation \citep[$e.g.$][]{Martig09, Phillips14}. 

The general consensus is that a complex mixture of both galaxy mass (mass quenching - here we define as quenching processes internal to the galaxy) and local environmental effects (environmental quenching - which we define as externally driven quenching processes) are the two key factors in driving this transition. However, the relative contributions for both effects as a function of other galaxy properties, such as stellar mass, are far from clear. Multiple studies have suggested that at high stellar masses (log$_{10}$[M$_{*}$/M$_{\odot}$]$>$10) galaxies are predominately quenched via non-environmental effects, which are proportional to their stellar mass and largely independent of local galaxy-galaxy interactions \citep[passive fractions increase to higher stellar masses, $e.g.$][]{Peng10,Wetzel13}. This is likely due to the increasing central to satellite fraction with increasing stellar mass. 

At intermediate stellar masses ($8.0<$log$_{10}$[M$_{*}$/M$_{\odot}$]$<$10.0) passive fractions are lowest \citep{Weisz15} and quenching likely occurs via starvation \citep[quenching timescales are comparable to gas depletion timescales;][]{Fillingham15,Peng15}, and at the lowest stellar masses (primarily Local Group galaxies at log$_{10}$[M$_{*}$/M$_{\odot}$]$<$8.0) quenching is likely to occur on much shorter timescales which are comparable to the dynamical timescale of the host halo, and are likely due to ram pressure stripping \citep{Fillingham15,Weisz15}. 

These studies suggest that different quenching processes dominate the transition from star forming to passive systems at distinct stellar masses, and that there is a characteristic stellar mass scale between environmental quenching via ram pressure stripping of cold gas, environmental quenching via starvation, and mass quenching - with these transitions occurring around log$_{10}$[M$_{*}$/M$_{\odot}$]$\sim$8 and log$_{10}$[M$_{*}$/M$_{\odot}$]$\sim$10 respectively. Analysis of these processes is made more problematic by varying gas-to-stellar mass ratios, which are found to increase significantly to lower stellar masses \cite[$e.g.$][]{Kannappan09} - typically  galaxies at these masses are abundant in star-forming gas, prior to being acted on by external processes. Note that historically there has been significant debate regarding the significance of ram pressure stripping in quenching dwarf galaxies in cluster environments \citep[$e.g.$][]{Davies88, Drinkwater01, Grebel03} - with the most recent predictions suggesting it is a necessary processes to produce the observed distribution of dwarf-spheroidal/elliptical galaxies.     

However, to date the majority of studies into the environmental effects of star formation are at the higher mass end of these characteristic masses (log$_{10}$[M$_{*}$/M$_{\odot}$]$>$9.5). The bulk of this analysis comes from the Sloan Digital Sky Survey \cite[SDSS, $e.g.$][]{Ahn14} and Galaxy And Mass Assembly survey (GAMA - see following section). In the former, the work of \cite{Peng10} and others, has highlighted the large scale environmental effects on star-formation in galaxies, while the extensive work of the Galaxy Pairs in SDSS group \citep[][]{Ellison08,Ellison10,Patton11,Scudder12} have probed deep into the effects of local galaxy-galaxy interactions on star-formation processes. In the latter, the GAMA survey has been used in numerous studies to investigate the local and large-scale environmental affects on star-formation \cite[$e.g.$][]{Wijesinghe12,Alpaslan15,Taylor15}. 

More recently, \cite{Davies15} has shown that the effects of galaxy-galaxy interactions in modifying star formation in GAMA galaxies is a complex process, where both pair mass ratio and primary/secondary status within the pair (central/satellite) determine whether a galaxy has its star formation enhanced or suppressed in the interaction. The Davies et al. work only covers the 9.5$<$log$_{10}$[M$_{*}$/M$_{\odot}$]$<$11.0 range, but directly shows that secondary galaxies in minor mergers (the satellite systems identified in the work discussed above) are the most likely to have their star-formation suppressed by an interaction, and that this suppression is occurring on relatively short timescales ($<100$\,Myr). As such, we may be directly witnessing starvation occurring in these galaxies as they transition from star-forming to quiescent systems. In contrast, Davies et al also find that enhancements to star formation are strongest in major mergers at around log$_{10}$[M$_{*}$/M$_{\odot}$]$\sim10.5$, which is close to the lowest passive fraction of satellites \citep[see summary in Figure 6 of][]{Weisz15}. Hence, it is likely that galaxy interactions at these masses are enhancing  star formation in satellite galaxies and reducing satellite passive fractions. These variations in star-formation processes are closely linked to interaction timescale, which is shortest around log$_{10}$[M$_{*}$/M$_{\odot}$]$\sim10.5$ and increases to lower stellar masses \citep[$e.g.$][]{Kitzbichler08}. If interacting galaxies take significant time to become passive \citep[$e.g.$ via starvation,][]{Fillingham15} then we may only witness the passive phase in low-mass systems (galaxies around log$_{10}$[M$_{*}$/M$_{\odot}$]$\sim10.5$ will merge too quickly to reach their passive phase). We explore this premise further in this paper.

In the lower stellar mass regime (log$_{10}$[M$_{*}$/M$_{\odot}$]$<$9.5) sample sizes are small and/or low mass pairs are not split from the global population ($e.g.$ the Galaxy Pairs in SDSS work). The \cite{Peng10} work does extend to log$_{10}$[M$_{*}$/M$_{\odot}$]$>$9.0 and they find that passive galaxies at the low-mass end of their sample predominately reside in over-dense environments, and such galaxies are largely non-existent in the field. They go on to argue that the mass-quenching efficiency for such low-mass galaxies is extremely low, and environmental quenching must be the dominant factor in producing red low-mass systems \cite[][]{Peng12}.

Some recent work has specifically targeted low mass galaxy-galaxy interactions - at around the transition mass between the proposed starvation quenching and ram pressure stripping quenching scenarios. These studies are also largely based on small nearby galaxy samples from SDSS \cite[$e.g.$][]{Geha12,Wheeler14} or Local Group galaxies \cite[$e.g.$][]{Weisz15} - which may be atypical of the general galaxy population. However, the results agree that galaxy-galaxy interactions are significant in forming passive low-mass galaxies, and a combination of these results suggest \cite[see Figure 6 of][]{Weisz15} that we are beginning to probe the characteristic mass scales between starvation and ram pressure stripping scenarios. 

In a converse, but complementary approach, \cite{Geha12} find that essentially all local log$_{10}$[M$_{*}$/M$_{\odot}$]$<$9.0 field galaxies are star forming, and that the passive fraction at these stellar masses drops rapidly as a function of distance from massive host galaxies. This also suggests that local environment plays a significant role in producing passive, low-mass systems \cite[also see earlier work of][etc]{Baldry06,Haines07}, and that isolated evolution alone fails to produce passive low mass galaxies.  

However, the environmental quenching of low-mass satellites is far from ubiquitous. Multiple recent studies \citep[$e.g.$][]{Phillips14,Wheeler14,Phillips15} have shown relatively modest fractions of $8.5<$log$_{10}$[M$_{*}$/M$_{\odot}$]$<$9.5 satellites are quenched (20-30\%) and that the efficiency of satellite quenching is deceasing with time \citep{Tinker13}. As such, quenching via starvation is likely to be driven by factors other than satellite stellar mass, potentially both host halo mass and interaction timescale.   \\

Theoretical modelling can offer some insights into this process, with contemporary semi-analytic simulations accurately reproducing the observed galaxy merger rates at low redshift \citep{Kitzbichler08}. These simulations also predict that secular transitions from active to passive galaxies are extremely problematic at the lowest stellar masses \citep[log$_{10}$(M$_{*}$/M$_{\odot}$)$<$9.0, they predict extremely low passive fractions in field environments, $e.g.$][]{Wheeler14}, and constrain the likely quenching timescale of interactions - highlighting that these timescales must be comparable to gas depletion timescales for intermediate mass satellites \citep[$e.g.$][]{Fillingham15}. Moreover, the relatively simple model predictions of \cite{Wheeler14} find that in order to produce the observed passive satellite population, requires either very-long quenching timescales or environmental quenching triggers which are not well matched to satellite accretion.

Previously, the GAMA sample has not been used to specifically target quenching in low mass galaxies. However, its extension to $\sim1$dex lower in stellar mass than SDSS at distances $\sim$10-100\,Mpc - closer to representing a cosmic average, high completeness to close pairs \citep{Robotham14}, extensive group catalogue \citep{Robotham13}, and multiple SFR diagnostics \citep[see][]{Davies15}, allows us to investigate the passive fraction in both pair and group environment and evaluate the relative contribution of environmental effects in producing quiescent galaxies. In the following paper, we identify passive and star-forming galaxies in the GAMA sample and investigate the environmental distribution of passive galaxies as a function of stellar masses, specifically focusing on the mass range close to the proposed starvation/ram pressure stripping transition. While this work displays the distribution of passive/star forming galaxies over a broad range of stellar masses, we specifically focus on passive galaxies at the low-mass end (log$_{10}$[M$_{*}$/M$_{\odot}$]$<9.5$) and the effect of galaxy-galaxy interactions. For a complementary analysis of the larger scale environmental effects on star-formation in spiral galaxies at log$_{10}$[M$_{*}$/M$_{\odot}$]$>9.5$, we refer the reader to Grootes et al., in prep.  

In this work, we take a slightly different approach to previous studies and do not initially define sources as either a central or satellite galaxy, but consider all pair systems. \cite{Phillips14} find that satellite galaxies are much more likely to be quenched if their host galaxy is also passive. As such, the same environmental effects may drive passive evolution in both centrals and satellites ($i.e.$ is it purely the interaction that drives the quenching in both galaxies, or the large scale environment?). In addition, we also consider major mergers, where the definition of central/satellite is somewhat ambiguous. We do split our sample into the primary/secondary pair galaxy class defined in \cite{Davies15}, which can be considered a central/satellite separation for minor mergers. We also investigate a number of different environments (group, pair, group and not pair, isolated - see Section \ref{sec:data} for details) to evaluate the relative effect of each on forming passive galaxies. 

However, care must be taken. In this work we assume that if quenched fractions correlate with specific environmental metrics, that there is a causal relationship between quenching processes and interactions with the local environment. Recent studies have shown that using $age$ $distribution$ $matching$ between numerical simulations and SDSS galaxies, the colour distribution in over-dense environments can be reproduced via halo assembly bias \citep[][]{Gao07} alone \citep[$e.g.$][]{Hearin13}. Moreover, further work has used the `galactic conformity' results of \cite{Kauffmann13} to show that assembly bias will naturally produce increased passive fractions in over-dense environments in the complete absence of any post infall processes, such as the environmental quenching processes discussed above \citep[][]{Hearin15}. However, these works focus on higher mass galaxies than those discussed here (log$_{10}$[M$_{*}$/M$_{\odot}$]$>9.8$). In addition, the \cite{Kauffmann13} work only finds conformity in gas rich, star-forming galaxies at log$_{10}$[M$_{*}$/M$_{\odot}$]$>9.8$ - not quenched gas poor systems. We also note that the \cite{Kauffmann13} `galactic conformity' is consistent with our previous results in \cite{Davies15}, where when considering the full pair galaxy population, both primary and secondary galaxies (i.e. central and satellites) have their SF boosted to a similar degree.    

This paper is structured as follows: in Section 2 we outline the GAMA data set and potential sample biases in our analysis, in Section 3 we discuss how we define passive and star-forming systems in our sample, in Section 4 we investigate the passive and star forming fraction as a function of stellar mass in various environments, in Section 5 we discuss the satellite passive fraction and suggest that the observed distribution is likely due to increasing interaction timescales at lower stellar masses, and in Section 5.3 we outline a simplistic model which can reproduce the observed trends in our data.

Throughout this paper we use a standard $\Lambda$CDM cosmology with {H}$_{0}$\,=\,70\,kms$^{-1}$\,Mpc$^{-1}$, $\Omega_{\Lambda}$\,=\,0.7 and $\Omega_{M}$\,=\,0.3.

\begin{figure*}
\begin{center}
\includegraphics[scale=0.53]{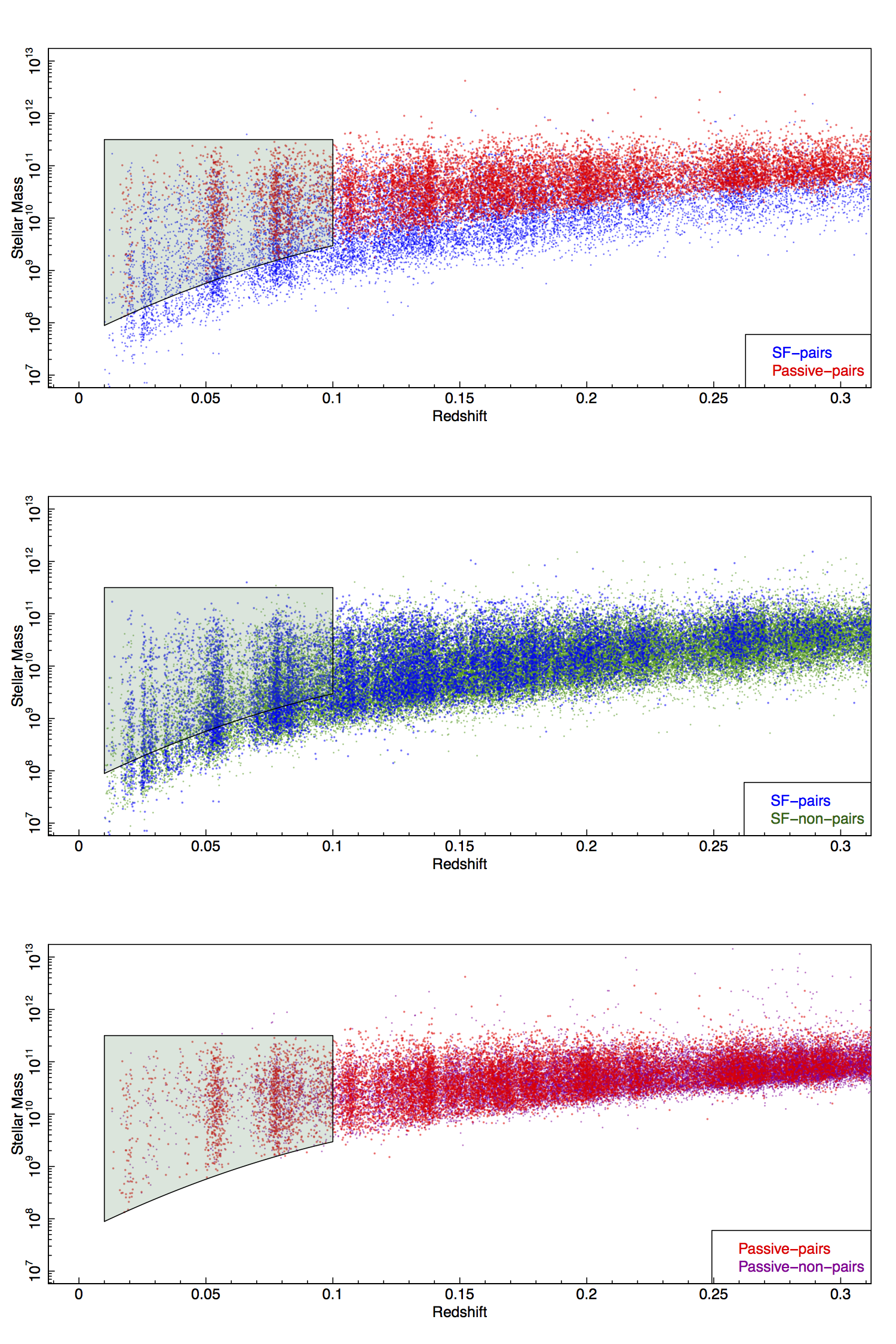}
\caption{The redshift-mass distribution of star-forming and passive galaxies (defined by H$\alpha$), split by those identified as being in pairs, and those not in pairs. The top panel shows the stellar masses of star-forming and passive pair galaxies as a function of redshift, highlighting the different selection limits for each class of galaxy - which potentially leads to biased pair classifications (see text in Section 2.1 for details). The middle panel show the comparison between star-forming galaxies in pairs and not in pairs, and the bottom panel, passive galaxies in pairs and not in pairs. The middle and bottom panel highlight that there is no significant difference in selecting star-forming or passive galaxies in pairs and non-pair environments. Our volume limited sample is displayed by the shaded polygon, which cover the $0.01<z<0.1$ and has a stellar mass limit as a function redshift given by Equation 1. We exclude all sources outside of this region and only include pair systems for which both galaxies lie within our volume limits - to exclude pairs which would not be identified as such if they had different passive/SF classifications (once again see text in Section 2.1 for details).}
\label{fig:mass_vs_z}
\end{center}
\end{figure*}

\section{Data}
\label{sec:data}

The data used in this work is derived and selected in a similar manner to that described in Davies et al. 2015. Here we briefly discuss the datasets used and refer the reader to previous work for further details.

The GAMA survey is a highly complete multi-wavelength database \citep{Driver11} and galaxy redshift ($z$) survey \citep{Baldry10,Hopkins13} covering 280\,deg$^{2}$ to a main survey limit of $r_{\mathrm{AB}}<19.8$\,mag in three equatorial (G09, G12 and G15) and two southern (G02 and G23) regions. The spectroscopic survey was undertaken using the AAOmega fibre-fed spectrograph \citep[][]{Sharp06,Saunders04} in conjunction with the Two-degree Field \citep[2dF,][]{Lewis02} positioner on the Anglo-Australian telescope and obtained redshifts for $\sim$250,000 targets covering $0<z\lesssim0.5$ with a median redshift of $z\sim0.2$, and highly uniform spatial completeness \citep{Robotham10,Driver11}. Full details of the GAMA survey can be found in \citet{Driver11} and \citet{Liske15}. In this work we utilise the first 5 years of data obtained and frozen for internal team use, referred to as GAMA II. The GAMAII is not currently publicly available, but is due to be fully released in the near future.    

In this work we shall use galaxy group and pair membership to delineate its local environment. We use the GAMA G$^{3}$C catalogue which includes the identification of all galaxy groups and pairs \citep[][also see \citealp{Robotham12,Robotham13,Robotham14, Davies15}]{Robotham11}. Briefly, the GAMA group catalogue is produced using a bespoke friends-of-friends based grouping algorithm which was tested extensively on mock GAMA galaxy light cones and assigns $\sim40\%$ of GAMA galaxies to multiplicity $>1$ pairs and groups \citep[][]{Robotham11}. In this work we define a group as a system with multiplicity N$>$2. The GAMA pair catalogue is further detailed in \cite{Robotham14} and \cite{Davies15}, where pair galaxies are selected on both radial velocity and physical positional offset using SDSS imaging and GAMA AAT spectra. In this work we use the full pair sample, defined as galaxies with physical separation $<$100\,h$^{-1}$\,kpc and velocity separation $<$ 1000 km\,s$^{-1}$ - over which pair galaxies have been found to affect each other's star-formation \cite[$e.g.$][]{Scudder12, Davies15}. While we can't rule out that pair galaxies are not truly interacting systems, but simply chance position and line of sight velocity correlated sources, care is taken in \cite{Robotham14} to determine the likelihood of pair systems being physically interacting. In this work, all pair systems are visually classified for observed disturbance using SDSS optical imaging. Figure 6 of \cite{Robotham14} displays the fraction of morphologically disturbed sources as a function of tangential and radial separation, and highlights that the pair selection method in GAMA is robust at identifying visually disturbed (definitively interacting) sources, specifically as small separations. Using the group and pair classifications, we also define samples of systems in pairs and/or groups, isolated galaxies which are not identified as being in a group or pair, and `locally isolated group galaxies' - which are in a group but not in a pair.     

Stellar masses for all galaxies are derived from the $ugriZJH$ photometry for all GAMA II galaxies \citep{Taylor11}. These stellar masses are calculated using a \cite{Chabrier03} -like IMF. As in Davies et al (2015) we further split pair galaxies into the primary or secondary systems (by stellar mass) and by pair mass ratio to identify pair galaxies as being either part of a potential major merger (mass ratio $<$3:1) or minor merger (mass ratio $>$3:1) - assuming the pair galaxies will eventually merge, which will clearly not be the case for all systems. However, in this work we use the well known terms of `major' and `minor' merger to denote pairs of different mass ratio. We restrict our sample to galaxies at $0.01<z<0.1$ and with GAMA redshift quality flag $>2$, for which all galaxies have pair/group assignments - the extent of the G$^{3}$C catalogue is $0.01<z<0.5$. We also minimise Active Galactic Nuclei (AGN) contamination in our sample, which may potentially bias SFR estimates, by excluding both optically bright AGN and composite sources using the BPT diagnostic diagram \citep[][]{Bladwin81} in an identical manner as in Figure 1 of \cite{Davies15}.

\begin{figure*}
\begin{center}
\includegraphics[scale=0.65]{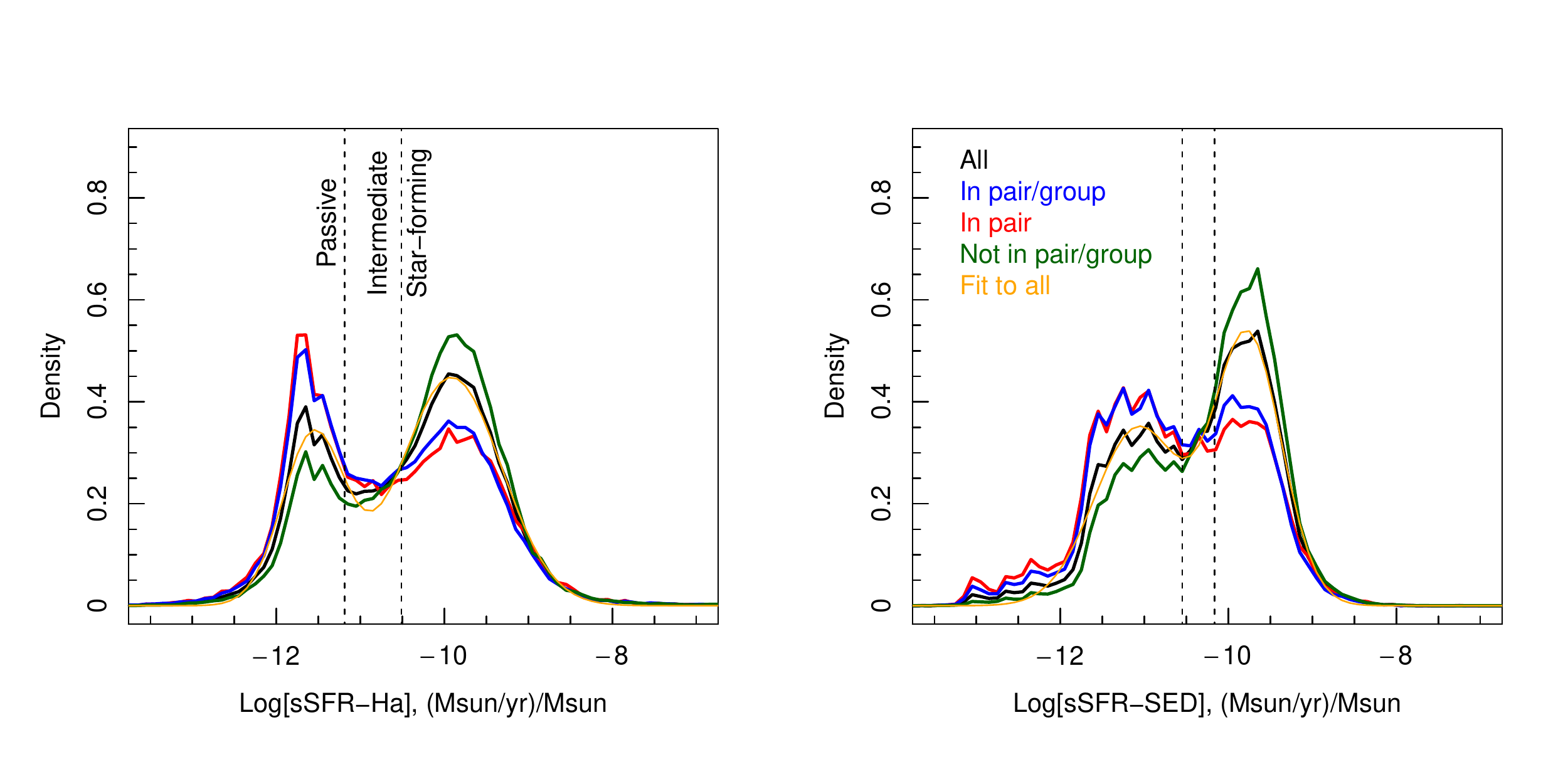}
\caption{Specific SFR distribution of galaxies in different local environments. Left: for H$\alpha$ derived SFRs and right: for \textsc{magphys} derived SFRs. {Colours show the different environmental classifications defined in Section 2}. Orange line displays the double gaussian least squares fit to the distribution of all galaxies (black line). Dashed vertical lines show the lower peak + 1$\sigma$ and the upper peak - 1$\sigma$ - of the orange line. We define galaxies as either passive (lower than left-most dashed vertical line), star-forming (higher than right-most dashed vertical line) or intermediate (between the two dashed vertical lines). We deem the bulk of galaxies in our passive/star-forming selection to be truly passive/star-forming, and ignore all intermediate sources in our subsequent analysis. Considering the different environments, passive galaxies are more likely to be in pair and/or group environments, whereas star-forming galaxies are more likely to be isolated ($i.e.$. the blue and red passive peak is larger than the full, black, distribution, while the green SF peak is larger than the full, black, distribution).}
\label{fig:sSFR_dist}
\end{center}
\end{figure*}

\begin{figure*}
\begin{center}
\includegraphics[scale=0.65]{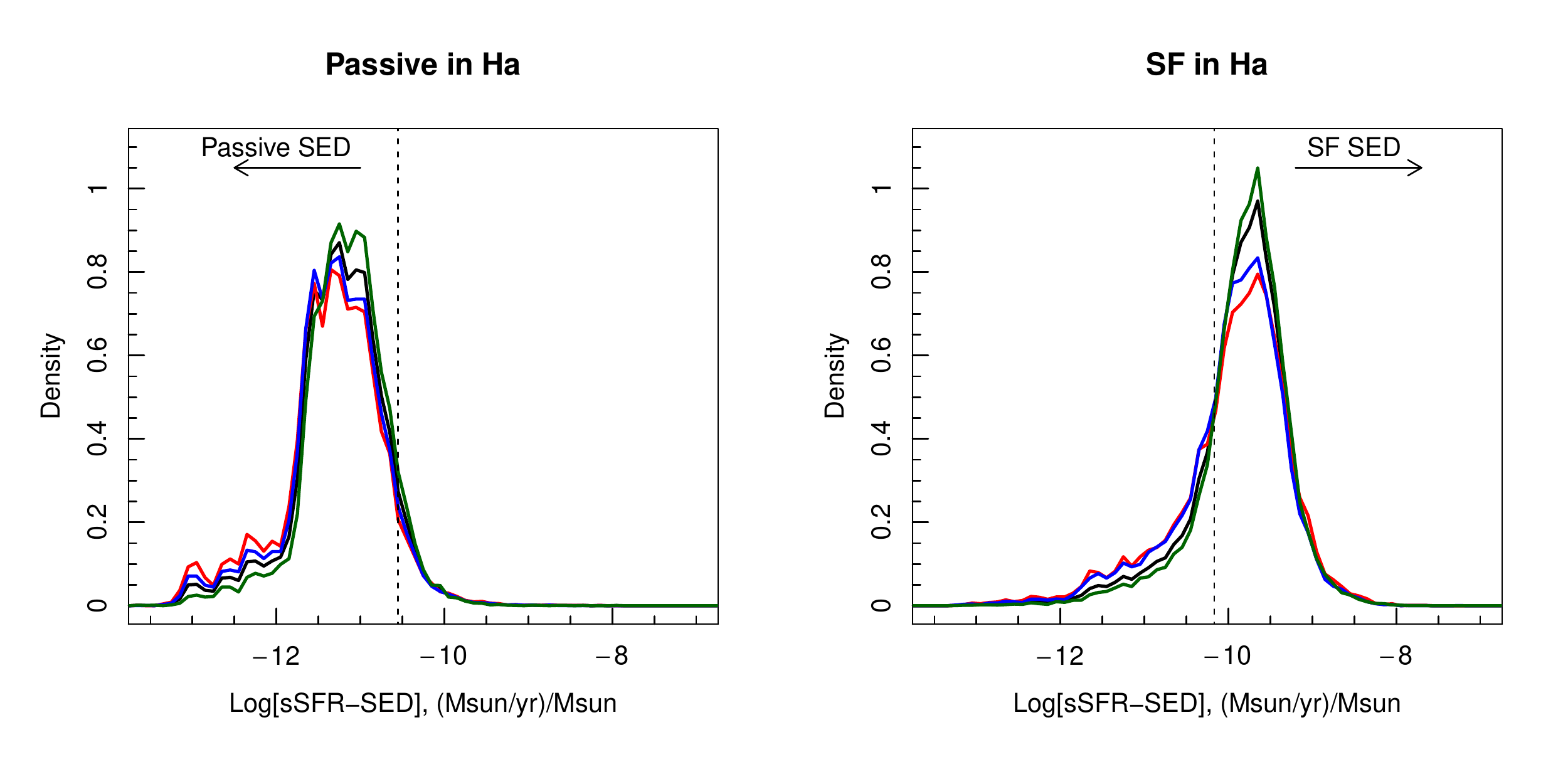}
\caption{The distribution of \textsc{magphys} sSFRs, split by sources selected as either passive (left) or star-forming (right) by their H$\alpha$ derived SFRs (left panel of Figure \ref{fig:sSFR_dist}). Vertical lines show the passive/intermediate/star-forming split of the right panel of Figure \ref{fig:sSFR_dist}. Clearly some sources defined as passive in H$\alpha$ are not when using \textsc{magphys} and vice versa. As such, we define an additional sample which are selected as star-forming or passive using $both$ classifications ($i.e.$ they are selected star-forming/passive in both H$\alpha$ and \textsc{magphys}). }
\label{fig:sSFR_dist_ha_sel}
\end{center}
\end{figure*}

\subsection{Potential Selection Biases}

Using the full GAMA data set directly is likely to induce significant biases in our results due to sample selection. Any spectroscopic sample will preferentially identify actively star-forming galaxies over passive galaxies, and this effect will be more apparent with increasing redshift. In order to minimise this, we define a volume limited sample based on the stellar mass detection limits of passive galaxies. Firstly we apply a maximum upper redshift limit of $z<0.1$ to minimise redshift evolution over our sample and biasing it towards high mass systems. Below $z\sim0.1$ we use a rolling sample selection at each stellar mass, defined by the lowest detection limit of the passive systems in GAMA (see the following section for SF/passive selections). To define this limit we split the passive population in $\Delta z=0.01$ bins from $0.01<z<0.2$ and determined the lower stellar mass limit which encompass the mean minus 2.25 $\times$ the standard deviation in each bin (found to accurately constrain the distribution of passive sources). We then fit a second order polynomial to the lower limits with the form:   

 \begin{equation}
log_{10}[M_{*}/M_{\odot}] > -65.2z^2+24.13z+7.71.
\end{equation}    

Figure \ref{fig:mass_vs_z} shows the volume limited sample used in this work. In this manner, we exclude star-forming systems which would not be detected in our sample if they were in fact passive and at the same stellar mass, and do not bias our selection based on our ability to identify lower-mass star forming systems. 

Potential biases in our analysis arise though group/pair assignments discussed above. There is a complex function which defines if systems are assigned to groups and pairs based on redshift, stellar mass, group mass, group occupation, pair mass ratio, etc. For example, we can not rule out that non-pair galaxies in GAMA have a faint companion which sits below the GAMA detection limits, or in fact that non-group galaxies do not have a significant number of sources below this limit (which if detected would cause the system to be identified as a group galaxy). 

However, for the former case, this scenario would place the galaxy in question as the central system in a low-mass major merger - where the bulk of this work will focus on satellite galaxies. In the latter, this scenario would place the system as the central galaxy in a low-mass group \cite[M$_{DM}\lesssim $ few$ \times 10^{12}$M$_{\odot}$,][]{Robotham11} - potentially where group environmental effects are less significant. At some point we must define a group/non-group selection at a specific group mass limit, as ultimately if selections were indefinitely extended to lower and lower stellar mass, almost all systems would reside in a group structure. As such, our group classification is defined as the systems falling within a group of M$_{DM}\gtrsim $ few$ \times 10^{12}$M$_{\odot}$ - and any results derived in this paper are conditional on this caveat.

Another potential bias arises from the dependance of the identification of pair systems based on the star-forming characteristics of each pair galaxy. Given the differing selection limits for passive and star forming galaxies, pair systems are more likely to be identified if both systems are star-forming, and additionally, passive + star forming pair systems are more likely to be identified if the larger mass galaxy is passive, than if the lower mass galaxy is passive. In order to minimise this bias, we have only included pair galaxies where $both$ systems meet our selection limits (as shown in Figure \ref{fig:mass_vs_z}). This essentially removes pair systems where one of the galaxies is star-forming, but lies below the selection boundary for passive system. We note that the galaxies in these pairs are completely removed from our analysis - $i.e.$ not included in our non-pair samples.

\section{Defining Star-Forming/Passive Systems}

To determine the affect of local environment on turning galaxies passive, we must first identify passive/star-forming systems in the GAMA dataset. We do this for two separate SFR indicators and both indicators in combination.  Firstly, we derive H$\alpha$ SFRs as in Davies et al (2015) using the most recent GAMA emission line measurements and the SFR calibration of \cite{Gunawardhana11}, based on the Kennicutt relation \citep{Kennicutt98} - with improved stellar absorption correction from \cite{Hopkins13}. The \cite{Gunawardhana11} calibration uses the Balmer decrement and GAMA aperture correction to convert observed H$\alpha$ fluxes to SFRs - for further details see Davies et al (2015). The GAMA emission line analysis assigns a H$\alpha$ measurement to all sources (irrespective of the presence of a detectable H$\alpha$ line) and as such, systems with no detectable H$\alpha$, will have low (but non-zero) SFRs. 

Secondly, we use the full SED fits to the 21-band photometric data available to all GAMA II sources \citep[GALEX-UV to $Herschel$-500$\mu$m,][]{Driver15b} using the energy balance code - \textsc{magphys}  \citep{daCunha08}. Full details of the GAMA \textsc{magphys} analysis will be presented in Wright et al. (in prep). 

Note that the H$\alpha$ SFRs described above use Salpeter-like IMFs, while our stellar masses and SED-based SFRs use Chabrier-like IMFs. As such, we scale the H$\alpha$ SFRs by a factor of 1.5 to account for this discrepancy \citep[][]{Dave08,Driver13}. However, our SF/passive selections used in this paper are based solely on the sSFR in our sample, and therefore are not sensitive to choice of IMF ($i.e.$ given a different IMF, we would simply find a different SF/passive selection).  

In \cite{Davies15} we use multiple SFR indicators to probe SF in pair systems (including MIR and FIR). However, in our current analysis we only use the H$\alpha$ and \textsc{magphys} SFRs. The MIR and FIR data used in GAMA have a significantly large PSF which may cause contamination between closely separated pair galaxies \citep[see discussion in][]{Davies15}. While care is taken in the GAMA to deblend flux to the relevant galaxies, this problem become increasingly difficult at low stellar mass, high mass ratio pairs (those primarily of interest in this work). As such, errors in deriving SFR in the MIR/FIR for low mass pair galaxies may be significant. While this error may also cause erroneous \textsc{magphys} fitting results, the energy balance fitting method and application of realistic galaxy templates, goes some way to reduce this issue. In addition, we will only use the \textsc{magphys} SFRs to remove sources from our passive and star-forming H$\alpha$ selection to produce a more robust sample (see below), and do not use the SFRs directly in the rest of the work. The H$\alpha$ SFR measurements are based on $3^{\prime\prime}$ aperture fibre measurements and as such, are not significantly affected by close source confusion. In this manner, we minimise any issues with deblending flux in closely separated sources.            

We note that the SFR timescales probed by the H$\alpha$ and \textsc{magphys} SFRs are different but comparable. H$\alpha$ probes star-formation on timescales of $\lesssim10$\,Myr, while the full SED SFRs typically probe $\lesssim100$\,Myr \cite[see][for a more comprehensive description of SFR indicator timescales]{Davies15}. H$\alpha$ is therefore likely to be sensitive to short duration changes in SF - which may be induced by galaxy interactions, but is also sensitive to aperture corrections which may bias derived SFRs. Here we use H$\alpha$ both independently and in combination with the more robust, but less sensitive to short timescale variation, full SED SFR.         

We derive specific SFRs (sSFR) using each indicator and the stellar masses of \cite{Taylor11}, and plot the distribution of all GAMA II sources for both H$\alpha$ and \textsc{magphys} SED sSFRs (black line, Figure \ref{fig:sSFR_dist}). In Figure \ref{fig:sSFR_dist} we see a clear distinction between star-forming and passive galaxies forming two largely distinct gaussian-like distributions in log[sSFR] space. This separation is more clearly defined for H$\alpha$ than the SED SFRs as there is a clear binary split between passive and star-forming systems (the presence or lack of the H$\alpha$ emission line). We fit the full distribution of sources (black line) with a double peaked normal distribution (orange line), using a least squares fit, and define sources as passive if they lie below the lower peak plus 1$\sigma$, star-forming if they lie above the upper peak minus 1$\sigma$ and `intermediate' if they fall between these two cuts - we then exclude the `intermediate' sources (see Figure \ref{fig:sSFR_dist} for separation lines). This selection primarily identifies sources which are either definitively passive or star-forming and removes any ambiguous sources. While the sample will be incomplete to all passive/star-forming galaxies, it applies strict constraints on identifying truly passive and star-forming systems and is likely to have little contamination from the ambiguous `intermediate' sources.

As noted above, initially we perform this selection independently, however, some sources are better defined using H$\alpha$ and some better defined using a full SED fit, in addition measurement error may cause either SFR indicator to be spuriously high/low in star-formation (see Figure \ref{fig:sSFR_dist_ha_sel}). As such, we also define samples of star-forming and passive galaxies which are selected as such using $both$ methods - these sources are likely to have a low contamination from incorrectly assigned sources as their star-formation rates are defined in two completely distinct and separate manners. We have visually inspected all sources in our passive selection at log$_{10}$M$_{*}$$<$9.0 and find that none show evidence of H$\alpha$ emission, and hence are likely to be truly passive galaxies. We also visually inspected a subsample of sources in our star-forming selection at the same mass range and find all show evidence of H$\alpha$ emission.

In Figures \ref{fig:sSFR_dist} and \ref{fig:sSFR_dist_ha_sel} we also split our sample into those galaxies defined as being in a pair, either pair or group and all other sources (not in a pair or group - which we define as isolated in this work). A higher fraction of passive sources are found in group and pair environments than are isolated - the passive peak is significantly larger for group and pair galaxies than for the isolated galaxies. This highlights the well known trend of galaxy star-formation in group environments, for example see \cite{Haines07,Haines08,Peng10,Mahajan10,Mahajan15}.

\section{Passive fractions of galaxies as a function of stellar mass and local environment}

\begin{figure*}
\begin{center}
\includegraphics[scale=0.6]{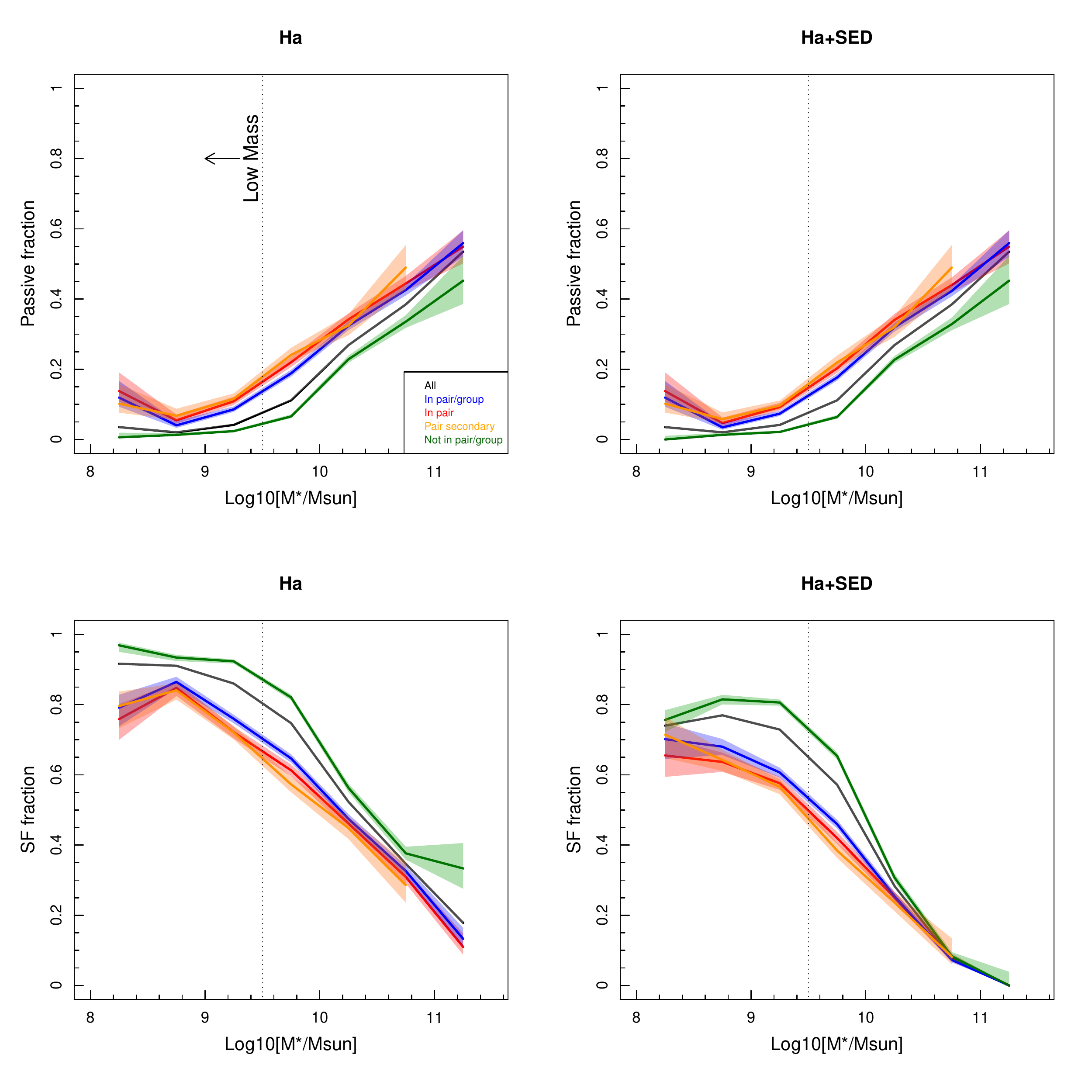}
\caption{Fraction of passive (top) and star-forming (bottom) galaxies (given the classifications in Figure \ref{fig:sSFR_dist}) as a function of stellar mass for H$\alpha$ and H$\alpha$+SED (\textsc{magphys}) derived SFRs. Translucent polygons display the 1$\sigma$ error range derived assuming a binomial distribution and colours represent different environmental metrics. The dashed vertical line displays the separation between high and low mass galaxies. To the left of this line are the stellar masses primarily discussed in the work. Groups (blue) and pairs (red) have higher passive fractions than the general galaxy population (black) at all stellar masses, echoing the results seen in Figure \ref{fig:sSFR_dist}. For H$\alpha$ and the joint classification there are essentially zero passive galaxies in non-pair/group environments at log$_{10}$M$_{*}$$<$9.5 - consistent with theoretical predictions. The orange line displays only the secondary galaxies in pairs (essentially satellites). The upturn at the lowest stellar masses in over-dense environments, potentially displays the increase in passive fractions from ram pressure stripping (see Section 4 for further details). Differences between the normalisation in SF fractions between H$\alpha$  and H$\alpha$+SED are due to the more stringent star-forming selection using both indicators.}
\label{fig:passive_fractions}
\end{center}
\end{figure*}

In this work we wish to investigate how local environment affects galaxies as a function of stellar mass, specifically at low stellar masses. In Figure \ref{fig:passive_fractions} we split our sample in $\Delta$M=0.5log$_{10}$(M$_{\odot}$) bins and show the fraction of galaxies defined as passive (top) and star-forming (bottom), in the previous Section, as a function of stellar mass and split by our environmental metrics - in a pair and or group, just in a pair, in pair and the secondary (satellite) galaxy and not in a pair or group. The translucent polygons in these and other plots display the 1$\sigma$ error range derived assuming a binomial distribution \citep{Cameron11}. 

The passive fraction decreases and star-forming fraction increases with decreasing stellar mass in all environments, as globally low-mass galaxies are more actively star-forming  \cite[the well known downsizing paradigm - that low-mass galaxies are more actively star-forming in the local Universe, $e.g.$][]{Cowie96}. However, it is interesting to note the differences between each environmental metric. Pair and group galaxies have a higher fraction of passive galaxies than the global distribution at all stellar masses (consistent with results seen in Figure \ref{fig:sSFR_dist}). We also isolate secondary galaxies in minor mergers (orange line) - these are essentially the satellite systems identified in previous work. While they have the highest low-mass passive fraction of all of our environmental metrics, this fraction is still lower than previous SDSS-based work. At 8.0$<$log$_{10}$[M$_{*}$/M$_{\odot}$]$<$9.0, \cite{Phillips14}, \cite{Geha12} and \cite{Wheeler14} find a passive fraction of $\sim20-30\%$ (albeit from the same sample), while we find $\sim$10-20\%. This difference could either be due to sample selection (our passive selection based on H$\alpha$ sSFR may be more stringent than their H$\alpha$ EW selection) or a difference in stellar mass estimations. For example, if we were to include all `intermediate' defined sources as passive in our H$\alpha$ selection, we would obtain a low-mass massive fraction of $\sim20-25\%$ (comparable to the previous results). We also note that our passive fractions at $10.0<$log$_{10}$[M$_{*}$/M$_{\odot}$]$<$11.0 are consistent with \cite{Phillips14}, \cite{Geha12,Wheeler14} and \cite{Wetzel13} at around 30-50\% \cite[see Figure 5 \& 6 of][for a summary of passive fractions as a function of stellar mass]{Weisz15}. Following this, we attribute the differences in normalisation of passive fractions between our current analysis and the previous SDSS-based work to be our sample selection methods, and our more stringent passive classification. From here on, we will continue to compare the overall trends in our results to the previous work, with the caveat that the overall normalisation in passive fractions differ.             

At the lowest stellar masses (log$_{10}$[M$_{*}$/M$_{\odot}$]$<9.0$) we find a very low passive fraction in non-group/pair environments for both of H$\alpha$ selection and H$\alpha$+SED SFR indicators combined ($<0.5\%$). This is consistent with \cite{Geha12} who find exceptionally low passive fractions for low-mass galaxies in isolated environments ($<0.1\%$). 

We also see an upturn in the passive fractions for pair/group galaxies in our lowest stellar mass bin. This is interesting considering the work on passive fractions of Local Group galaxies in \cite{Weisz15} in combination with passive galaxies at higher masses. As discussed previously, they find passive fractions decrease with decreasing stellar mass to log$_{10}$[M$_{*}$/M$_{\odot}$]$\sim9.0$M$_{\odot}$ and then rise again at log$_{10}$[M$_{*}$/M$_{\odot}$]$<8.0$M$_{\odot}$ for Local Group galaxies - suggesting this transitions represent changes between starvation quenching at intermediate masses, and ram pressure quenching at low-masses. This is consistent with the trends in our results and we may be witnessing the start of the low-mass upturn in passive fractions, potentially driven by the ram pressure stripping scenario. However, we find that this upturn occurs at slightly higher masses than in \cite{Weisz15} -  log$_{10}$[M$_{*}$/M$_{\odot}$]$<8.5$M$_{\odot}$. This is not surprising given that the transition point at these masses has not previously been well probed by either the Local Group or nearby galaxy studies, and is consistent with the shape of the satellite passive fraction at these masses seen by \cite{Wheeler14}.       

Considering the star-forming fractions, we see the well-studied and general trend of increasing star-forming fractions at low stellar masses, and that non-pair/group environments have a higher star-forming fraction than group/pair galaxies (specifically at the highest stellar masses), once again completely consistent with many previous studies \cite[$e.g.$][]{Peng10}. \\

\subsection{Environmental Passive Fraction}

Taking this analysis further, it is interesting to consider the \textit{relative} fraction of passive galaxies in different environments as a function of stellar mass ($i.e.$ of all passive galaxies, what fraction live in a pair, in a group, in a group and/or pair, in a group but not in a pair, or is isolated). We define this as the `environmental passive fraction'. In this manner we can compare and contrast the relative contributions of different environments in producing passive galaxies. For example, are there particular stellar masses where pair classification appears more dominant than group classification in producing passive galaxies? And/or are there stellar masses where local environment appears to have little effect on forming passive galaxies?
 
Figure \ref{fig:all_fractions} displays the fraction of all passive (top) and star-forming (bottom) galaxies as a function of environment and stellar mass. Note that the bottom panel is not the converse of the upper panel, as in our selections not all galaxies are defined as either passive or star-forming. As such, comparing the top and bottom panels directly is problematic. In essence one should consider the top panel as a measure of how environment turns galaxies passive and the bottom panel how environment boosts star formation. These two scenarios are subtly, but importantly distinct. 

Considering the passive fraction, at low stellar masses pair/group galaxies dominate the distribution. At log$_{10}$[M$_{*}$/M$_{\odot}$]$<9.5$, $>60\%$ of all passive galaxies are in a pair or group (blue), with pair classification being the dominant factor at log$_{10}$[M$_{*}$/M$_{\odot}$]$<9.0$. The fraction of passive galaxies in pairs drops with increasing stellar mass until it reaches a minimum at log$_{10}$[M$_{*}$/M$_{\odot}$]$\sim10.25$ and then rises (red). This is mirrored in the isolated passive galaxy fraction, which is small at low stellar masses, rises to log$_{10}$[M$_{*}$/M$_{\odot}$]$\sim10.25$ and then drops at higher stellar masses (green). Non-pair group galaxies (purple) also have a low passive fraction at low stellar masses and rises progressively to higher stellar masses. The full group distribution (which we remind the reader also contains pairs which are in groups) is relatively flat at $\sim0.5$ until log$_{10}$[M$_{*}$/M$_{\odot}$]$\sim10.25$ and then progressively rises to higher stellar masses. From this figure we identify four key observables, and inferences from them:\\

\noindent $\bullet$ \textbf{Pair galaxies dominate the environmental passive fraction at the lowest stellar masses, and the passive fraction increases  with decreasing stellar mass}. As such, local galaxy-galaxy interactions are likely to play a significant role in turning low-mass galaxies passive. There are very few passive non-pair galaxies at low-masses (either in the field or non-pair group galaxies), which is in agreement with predictions that low-mass galaxies do not become passive via isolated evolution alone, and is consistent with previous results.    \\

\noindent $\bullet$ \textbf{At log$_{10}$[M$_{*}$/M$_{\odot}$]$\sim10.25$ the environmental passive fraction in pairs is lowest}. This point is close to the peak in the major merger rate from \cite{Robotham14}. As shown in Davies et al (2015), in major mergers at this mass range both galaxies have their star-formation strongly enhanced by the interaction. As such, the passive fraction in pairs drops significantly, which is echoed in an increase in the passive fraction in isolated galaxies (it is likely that this peak is not driven by an increase in passive galaxies in the field, but a decrease in passive galaxies in pairs).  In fact, at these masses passive galaxies are more likely to be found in the field rather than in a pair and are almost equally likely to be in a group as the field. It appears that log$_{10}$[M$_{*}$/M$_{\odot}$]$\sim10.25$ is an important transition point where the dominant factor in producing passive galaxies changes from environmental to mass quenching (see discussion), and where we see the strongest enhancement of star-formation in interactions \citep[close to the peak in the major merger rate, where SF is most strongly enhanced, see][]{Davies15}.  \\

\noindent $\bullet$ \textbf{At higher stellar masses, passive galaxies are increasingly found in group environments}. This is displayed by the rise of environmental passive fractions in non-pair group galaxies (purple). Such systems are in a group environment but are not currently undergoing a local galaxy-galaxy interaction. This rise highlights that high-mass passive galaxies are more likely to be in groups, $but$ $not$ $necessarily$ in pairs ($i.e.$ it is the group environment/assignment that drives the passive classification, $not$ the local galaxy interactions). This suggests that as we move to higher stellar masses, mass quenching becomes more important at making galaxies passive. A high fraction of passive galaxies are still found in pairs, but there is also a high fraction in isolated galaxies. As such, the interaction is not likely to be the dominant factor in modifying SF at these masses.   \\

Looking  at this from the other side, the bottom panels of Figure \ref{fig:all_fractions}, show the fraction of star-forming galaxies in each environment as a function of stellar mass. At low stellar masses we consistently see that star-forming galaxies are more likely to be found in the field than in group/pairs. At log$_{10}$[M$_{*}$/M$_{\odot}$]$>$10.25 we see a large increase in the fraction of star-forming galaxies that reside in groups and pairs. This suggest that environmental effects do not have a significant impact on boosting the star-forming fraction in low-mass galaxies but the effect becomes increasingly important at higher stellar masses - which is consistent with the top panels. Once again it is interesting to note that these transitions occur around the log$_{10}$[M$_{*}$/M$_{\odot}$]$\sim$10.25 point.

\subsection{Comparison to Pair Classifications}

As alluded to above and discussed at length in Davies et al (2015), the galaxy assignment within pairs as either primary or secondary (central/satellite) and the pair mass ratio (major/minor merger) plays a significant role on the effects of the interaction on the SF properties of the galaxies. Note that the Davies et al work only considers galaxies in the 9.5$<$log$_{10}$[M$_{*}$/M$_{\odot}$]$<$11 range but finds that star-formation is enhanced in major mergers and the primary galaxies of minor mergers, but suppressed in the secondary galaxies in minor mergers (satellites).  As such, can we relate the dominant pair galaxy assignments discussed in Davies et al, as a function of stellar mass, to the fraction of passive/star-forming galaxies in pairs as a function of stellar mass, discussed above? For example, if low-mass galaxies are made passive by an interaction, the bulk of these galaxies should be secondary galaxies in minor mergers.  

Figure \ref{fig:all_fractions_split} displays the pair class assignments of Davies et al. as a function of stellar mass. At low stellar masses galaxies are predominantly the secondary galaxy in an interaction (lower mass - left panel) and in equally in major/minor mergers (middle panel). Combining these classifications in the right-most panel, we find that at low stellar masses the pair galaxy population is dominated by the secondary galaxies of minor mergers - those which Davies et al. find to have their star-formation suppressed. At around log$_{10}$[M$_{*}$/M$_{\odot}$]$\sim10$ the population transitions to being major-merger dominated and we see equal contributions from primary/secondary and major/minor classes. At log$_{10}$[M$_{*}$/M$_{\odot}$]$>$10 the distribution is strongly dominated by primary galaxies in minor mergers. These sources are likely to be large central galaxies, undergoing interactions with small satellites. 

\begin{figure*}
\begin{center}
\includegraphics[scale=0.6]{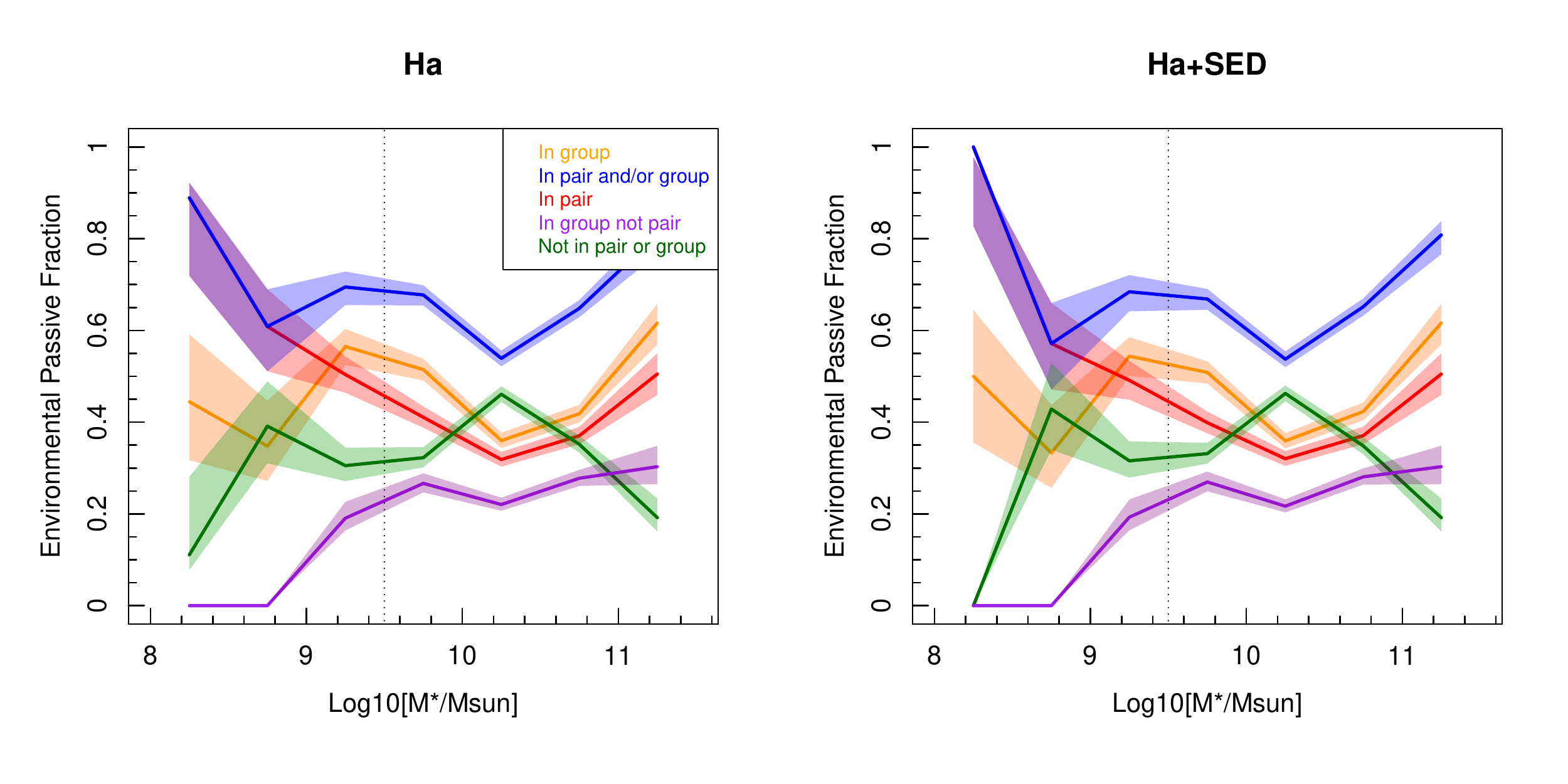}
\includegraphics[scale=0.6]{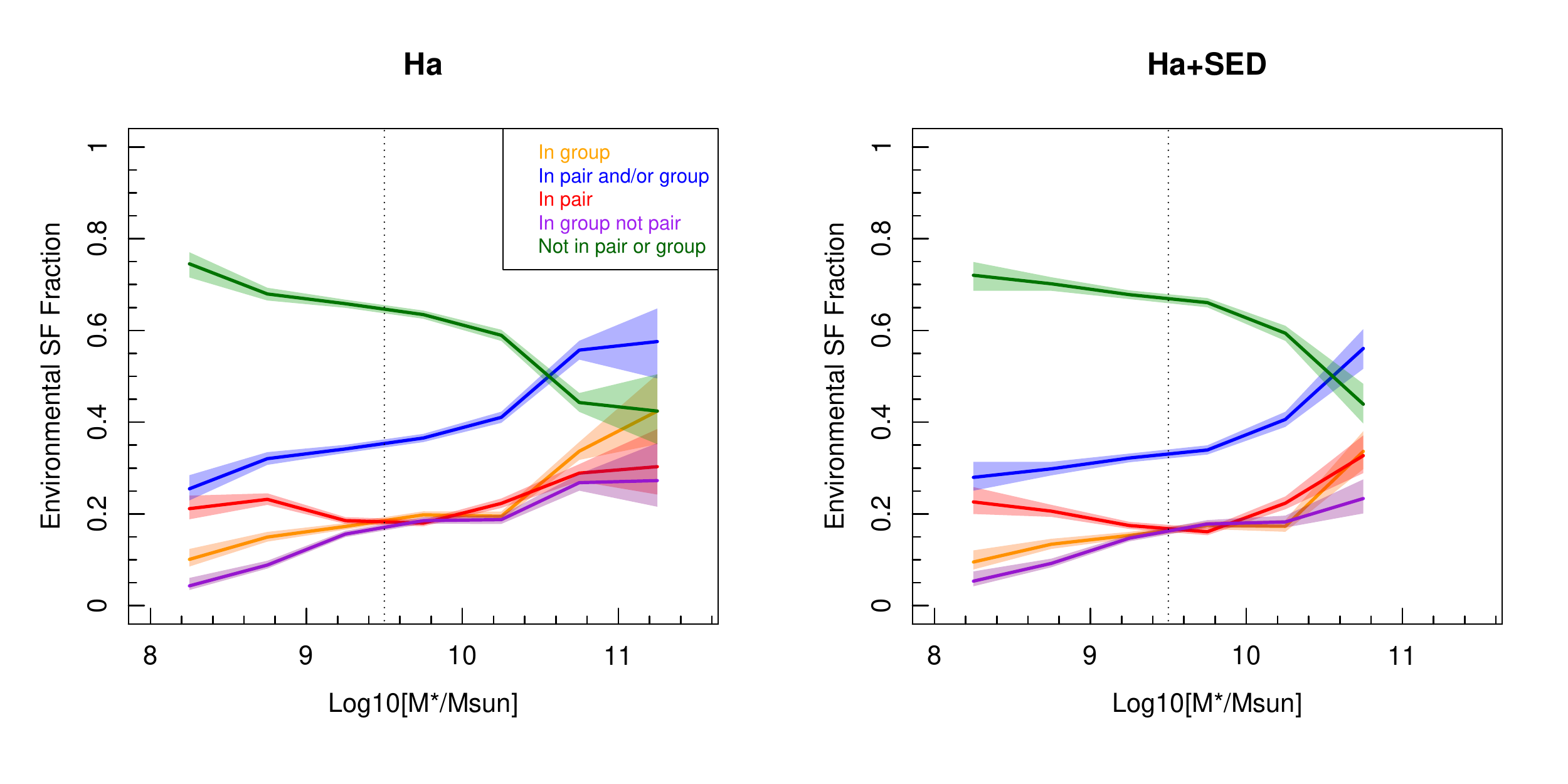}
\caption{The passive (top) and star-forming (bottom) fractions of galaxies in different local environments ($i.e.$ of all passive/star-forming galaxies with environmental metrics, which fraction fall into groups, pairs, or are isolated).  Once again, the translucent polygons display the 1$\sigma$ error range derived assuming a binomial distribution. In this figure we also show galaxies which are classified as being in a group, but not in a pair (purple line). For H$\alpha$ SFRs $>$60\% of all log$_{10}$M$_{*}$$\sim$9.5 passive galaxies live in pairs and/or groups - with pair classification being the most significant factor at the lowest stellar masses. Conversely a very small fraction of passive galaxies at this these stellar masses are isolated. The passive fraction of isolated galaxies increases with stellar mass until log$_{10}$M$_{*}$$\sim$10.5 (the point where the number of passive galaxies is equal in isolated and group environments) and then drops. The dashed vertical line once again displays the stellar mass at which we define `low-mass' systems (log$_{10}$[M$_{*}$/M$_{\odot}$]$<9.5$). }
\label{fig:all_fractions}
\end{center}
\end{figure*}

\begin{figure*}
\begin{center}
\includegraphics[scale=0.46]{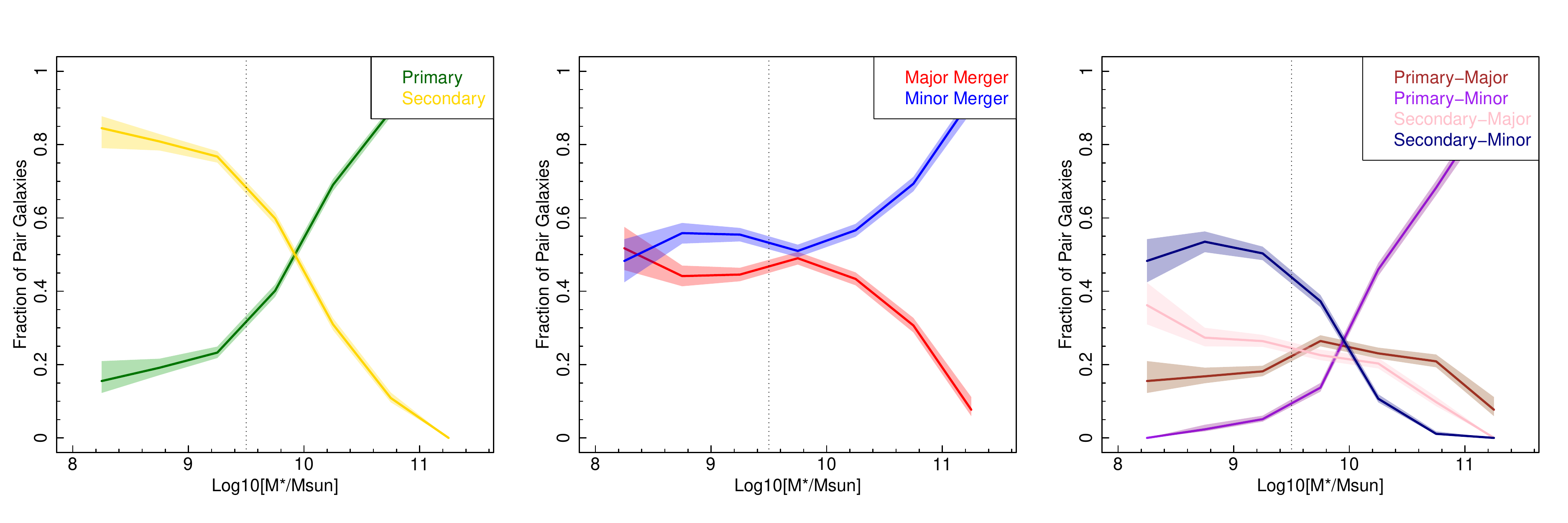}
\caption{The fraction of all pair galaxies as a function of stellar mass, split into pair classifications of primary/secondary pair status (central/satellite) and major and minor mergers (1:3 mass ratio). It is interesting to note that low-mass galaxies in pairs (log$_{10}$[M$_{*}$/M$_{\odot}$]$<9.5$M$_{\odot}$) are highly likely to be the secondary galaxy in a minor merger - this is exactly the regime where star-formation is found to be suppressed in \citet{Davies15}. Hence, we see the highest passive fractions in low-mass pairs, and this is driven by the fact that a higher fraction of these systems are the secondary galaxy in a minor merger. As such, the galaxy-galaxy interaction is likely to be the driving force in turning these galaxies passive. Once again, the translucent polygons display the 1$\sigma$ error range derived assuming a binomial distribution and the dashed vertical line displays the stellar mass at which we define `low-mass' systems. }
\label{fig:all_fractions_split}
\end{center}
\end{figure*}

There are a number of key observations in these figures that we can relate to the distributions in Figure \ref{fig:all_fractions}:\\

\noindent $\bullet$ At the lowest masses, secondary-minor galaxies dominate. In Davies et al. this is the only class of galaxy where star-formation is suppressed in the interaction. This is consistent with the high passive fraction/low star-forming fraction in pairs at these masses.   \\
 
\noindent $\bullet$  The mass at which the majority of galaxies are primaries of minor mergers (log$_{10}$[M$_{*}$/M$_{\odot}$]$>10.0$), and where secondary-minor galaxies are no longer the dominant class, is similar to the mass where the pair passive fraction is lowest. We know from Davies et al these pair classes are where star-formation is primarily enhanced in galaxy interactions, as such these results are consistent. \\

\noindent  $\bullet$ At the highest stellar masses the  primary galaxies in minor mergers dominate. These are likely to be central galaxies undergoing mergers with smaller satellites. The interaction is unlikely to strongly modify their star-formation processes. These systems may have also already become passive via mass quenching processes, and as such can not be affected by the interaction. This is consistent with the passive fractions at these masses not being strongly dependent on local galaxy interactions (as discussed previously).    \\
  
While the distributions in Figure \ref{fig:all_fractions_split} are not surprising (essentially showing that low-mass galaxies are likely to merge with larger galaxies, and high mass galaxies are likely to merge with smaller galaxies - as you would expect simply from number statistics and hierarchical formation) they do highlight the interesting transition points in stellar mass where different merger classes dominate. It is intriguing that the transition point in stellar mass where primary/secondary galaxies dominate, and where all merger classes are equally represented, is almost identical to the point where the passive fraction in pairs/groups and non-pairs/groups is the same,  log$_{10}$[M$_{*}$/M$_{\odot}$]$\sim10.25$ - potentially the transition point where environmental effects no longer have a strong impact in forming passive galaxies. This also suggests that the environmental quenching is only significant in secondary (satellite) galaxies and does not strongly affect primary (central) systems, potentially contradictory to the \cite{Phillips14} results of satellite/central joint quiescence.         
  
\subsection{Summary of passive galaxies as a function of stellar mass and environment}

\begin{figure}
\begin{center}
\includegraphics[scale=0.30]{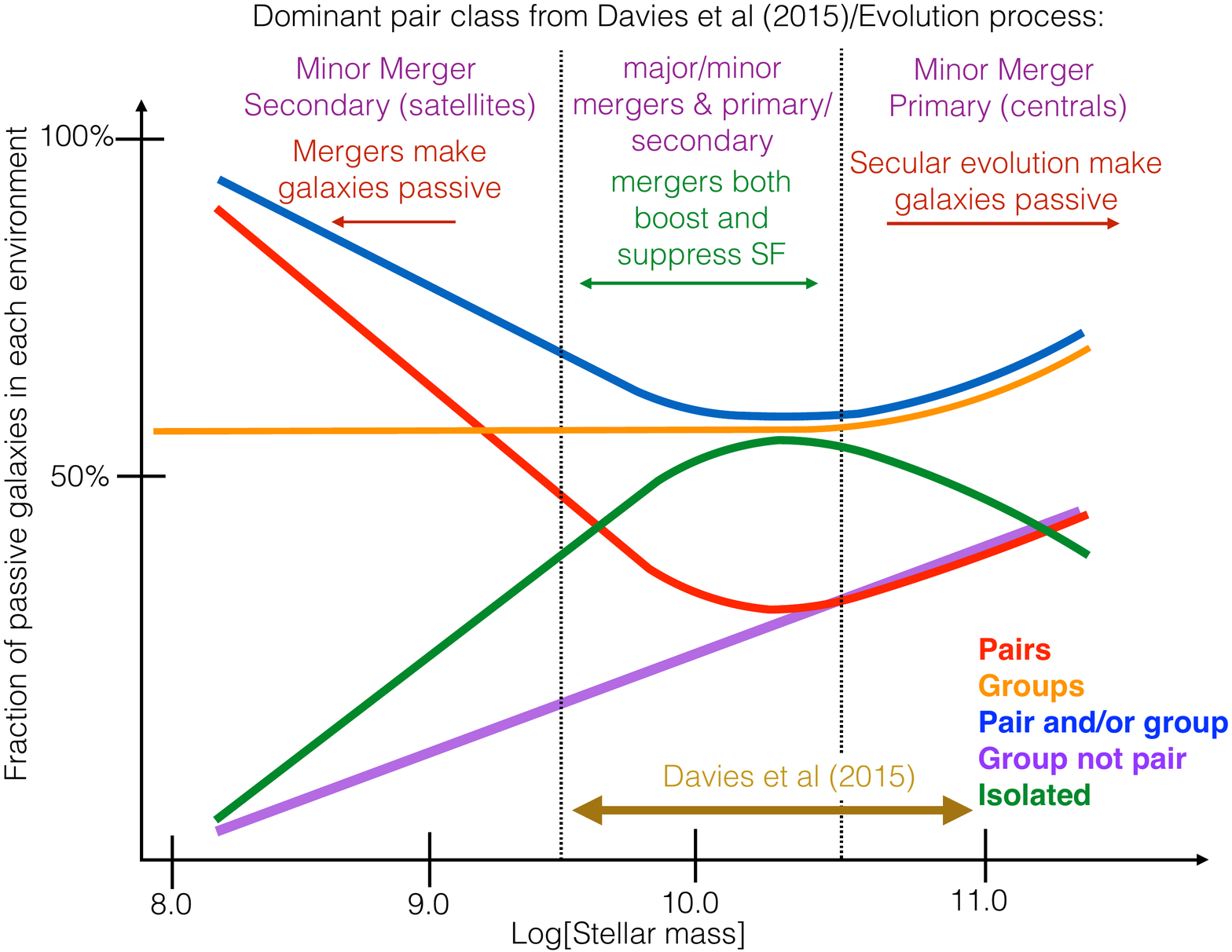}

\caption{Left: Cartoon representation of the top left panel of Figure \ref{fig:all_fractions} showing the environment in which passive galaxies reside as a function of stellar mass. Included in this figure are regions where various classes of pair galaxies dominate and how mergers/secular evolution drive galaxy evolution as a function of stellar mass. At low-masses the majority of galaxies are secondaries in major mergers. As such they have their SF suppressed and become passive. At higher masses galaxies are a mixture of different merger types and the fraction of passive galaxies is roughly equal between pairs and isolated systems. Here galaxies are made both passive and star-forming by interactions and are either star forming or passive in the field. At higher masses still, secular evolution takes over. Locally isolated group galaxies (in a group but not in a pair), field galaxies and pair galaxies all show similar passive fractions, suggesting environment has little effect on modifying star formation in these galaxies. At these masses pair galaxies are largely the primary galaxy in a minor merger and are not significantly affected by the interactions with significantly lower mass galaxies. }

\label{fig:cartoon}
\end{center}
\end{figure}

Clearly there is a complex interplay of galaxy interactions of different types and environment affecting the star-formation evolution of galaxies - where it is likely that pair status (primary/secondary), pair mass ratio and absolute stellar mass all govern how an interaction modifies star formation. Here we aim to piece together the observables discussed above to produce a self-consistent model of how interactions affect star formation in galaxies as a function of stellar mass. We use the key points outlined in the previous section to produce a cartoon representation of the local environmental effects on star formation as a function of stellar mass (Figure \ref{fig:cartoon}). In this proposed scenario low-mass galaxies ($8.0<$log$_{10}$[M$_{*}$/M$_{\odot}$]$<9.5$) are strongly affected by an interaction and are primarily made passive - with this effect increasing with decreasing stellar mass. In this regime the pair galaxy population is dominated by the secondary galaxies in minor mergers, which is consistent with the \cite{Davies15} analysis at higher stellar masses, which finds that this class of galaxies shows the strongest suppression during an interaction. 

At 9.5$<$log$_{10}$[M$_{*}$/M$_{\odot}$]$<10.5$ we see a mix of different interaction processes, pair galaxies are both suppressed and enhanced in star-formation depending on their pair mass ratio, as such galaxy interactions produce a broad range of effects. This range covers the peak major merger rate, where star-formation is most enhanced and therefore, pair passive fractions are lowest. We also note that this mass range is similar to the characteristic turnover mass in the metallicity-specific SFR (Z-sSFR) relation for all GAMA galaxies found in \cite{Lara-lopez13}. In this work, they also find high dispersion in their relation at this point and propose that varying gas mass and metallicity produce a mixing of processes. They predict that in high mass galaxies sSFR correlates with metallicity and in low mass galaxies an anti-correlation is observed. This is being explored further in upcoming work (Lara-L\'opez in prep).     

At log$_{10}$[M$_{*}$/M$_{\odot}$]$>10.5$ the passive fraction is increasingly dominated by non-pair group galaxies. We see roughly equal fractions of passive galaxies in pairs, non-pair group galaxies and non group/pair galaxies, suggesting galaxy interactions have little effect in turning galaxies passive - mass quenching is likely to be the dominant process driving star formation evolution. We do see an increase an the group passive fraction at these masses (as has been found for many previous studies). However, the rise of passive fractions in non-pair group galaxies suggests that this is due to factors other than local galaxy-galaxy interactions. Potentially group galaxies at these masses are significantly older than those in the field, and as such have had sufficient time to consume all of their star-forming gas. We also reminded the reader that the recent simulations of \cite{Hearin15} highlight that passive fractions at log$_{10}$[M$_{*}$/M$_{\odot}$]$>10.0$ can potentially be explained by assembly bias, and as such, may not need quenching processes to produce the observed passive fraction.   \\

\section{Discussion}

In the following subsections we further investigate the affect of galaxy-galaxy interactions on causing quiescence, focusing just on the satellite pair systems (which we now define as the lower mass galaxy in each pair irrespective of pair mass ratio - previously called `secondary'). In Subsection \ref{sec:pass_f} we investigate the satellite passive fraction as a function of stellar mass, to highlight the increasing effect of interactions at lower stellar masses, in Subsection \ref{sec:time} we relate these trends in this passive fraction to interaction timescale and highlight that our observed distribution can potentially be explained by varying interaction timescales as a function of stellar mass, and in Subsection \ref{sec:toy} we produce a simplistic model to test this premise.

\subsection{Satellite passive fraction}
\label{sec:pass_f}

Figure \ref{fig:summary} shows the satellite passive fraction (as defined by our H$\alpha$ SFR) of all pair galaxies scaled by the global passive fraction of all galaxies in our sample (top) and the passive fraction of all non-pair galaxies (middle). This scaling removes any dependance of the global galaxy passive fraction as a function of stellar mass and directly measures the increased chance of being passive while in a galaxy-galaxy interaction. In Figure \ref{fig:summary} a value of 1 indicates that the pair passive fraction at a given stellar mass is identical to the global/non-pair passive fraction, while values $>1$ highlight that interactions have some affect in making galaxies passive. At log$_{10}$[M$_{*}$/M$_{\odot}$]$>10.25$ (the transition mass discussed previously), galaxy interactions appear to have little affect on turning galaxies passive, as the satellite passive fraction is similar to the global/non-pair value. This is consistent with the recent work of \cite{Behroozi15}, who find only modest changes to passive fractions due to major mergers at log$_{10}$[M$_{*}$/M$_{\odot}$]$\sim10.0-10.5$. However, at lower stellar masses we see an increasingly strong trend of interacting systems having a larger passive fraction than the global/non-pair samples. This suggests that at low stellar masses interactions significantly suppress the star formation in galaxies. 

This is consistent with previous results at this mass regime \citep[][]{Phillips14,Phillips15,Wheeler14,Fillingham15}, but also shows the direct trend of the increasing effect of interactions in quenching galaxies as a function of decreasing stellar mass from log$_{10}$[M$_{*}$/M$_{\odot}$]$=11$ to log$_{10}$[M$_{*}$/M$_{\odot}$]$=8$. Clearly lower mass galaxies are more strongly affected by their very local environment than those at higher masses.   

\subsection{Interaction Timescales}
\label{sec:time}

Previous studies comparing passive galaxies to numerical simulations \cite[$e.g.$][]{Wheeler14,Fillingham15} suggest that at log$_{10}$[M$_{*}$/M$_{\odot}$]$>8$ significantly long quenching timescales are required to form passive galaxies via starvation. As such, it is interesting to consider whether the correlation between stellar mass and the effect of quenching in interactions is directly related to the interaction timescale. For example, if environment quenching occurs on long timescales, then galaxies that have been undergoing an interaction for sufficiently long time should show a high passive fraction. Conversely, short duration interactions may not be long enough for the galaxies to become passive. We remind the reader that H$\alpha$ SFRs probe timescales of $<0.01$\,Gyr and therefore are essentially a measure of the instantaneous SFR during the interaction.         

From numerous simulations \cite[$e.g.$][]{Boylan-Kolchin07,Wetzel15} we find that galaxy merger timescales are dependant on both the pair mass ratio and the host halo mass, where high pair mass ratio and low-halo mass leads to long duration mergers \citep[also see][- where interaction timescale goes as M$_{total}^{-0.3}$ for galaxies at log$_{10}$M$>9.5$]{Kitzbichler08}. To investigate the correlation between interaction timescale and the effect on pair passive fractions, we assume the most simplistic model, where:

\begin{equation*}
\tau_{merge} \propto \dfrac{M_{p}}{M_{s}} \times  M_{p}^{-1/2},
\end{equation*}

\noindent and $M_p$ and $M_s$ are the masses of the primary and secondary galaxies respectively and take $M_{p}^{-1/2}$ as a proxy for the primary galaxy's dynamical timescale at the viral radius, $(r_{vir}^3/GM_{host})^{1/2}$. This correlation essentially assumes a fixed orbital energy, angular momentum, host galaxy viral radius \cite[which is potentially appropriate at low stellar masses, where the size-mass relation flattens, see][]{Lange15} and mass to light ratio. We highlight that this analysis is speculative, but is only intended to show that to first order the distributions of dynamical timescales of interactions is similar in shape to the effect of interactions in turning galaxies passive. Given the limitation of the data available to us, we believe this approximation to be adequate.

\begin{figure*}
\begin{center}
\includegraphics[scale=0.5]{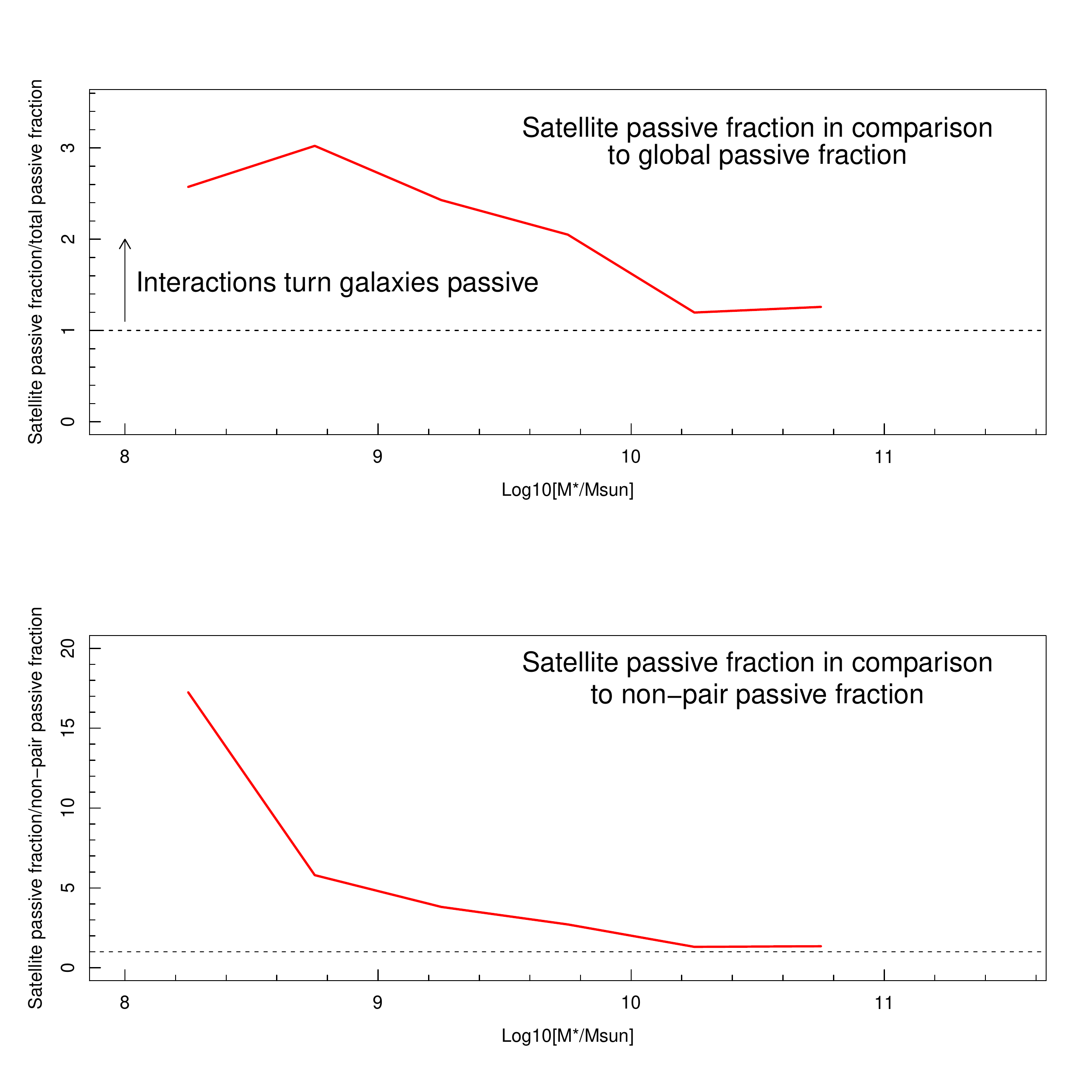}
\includegraphics[scale=0.5]{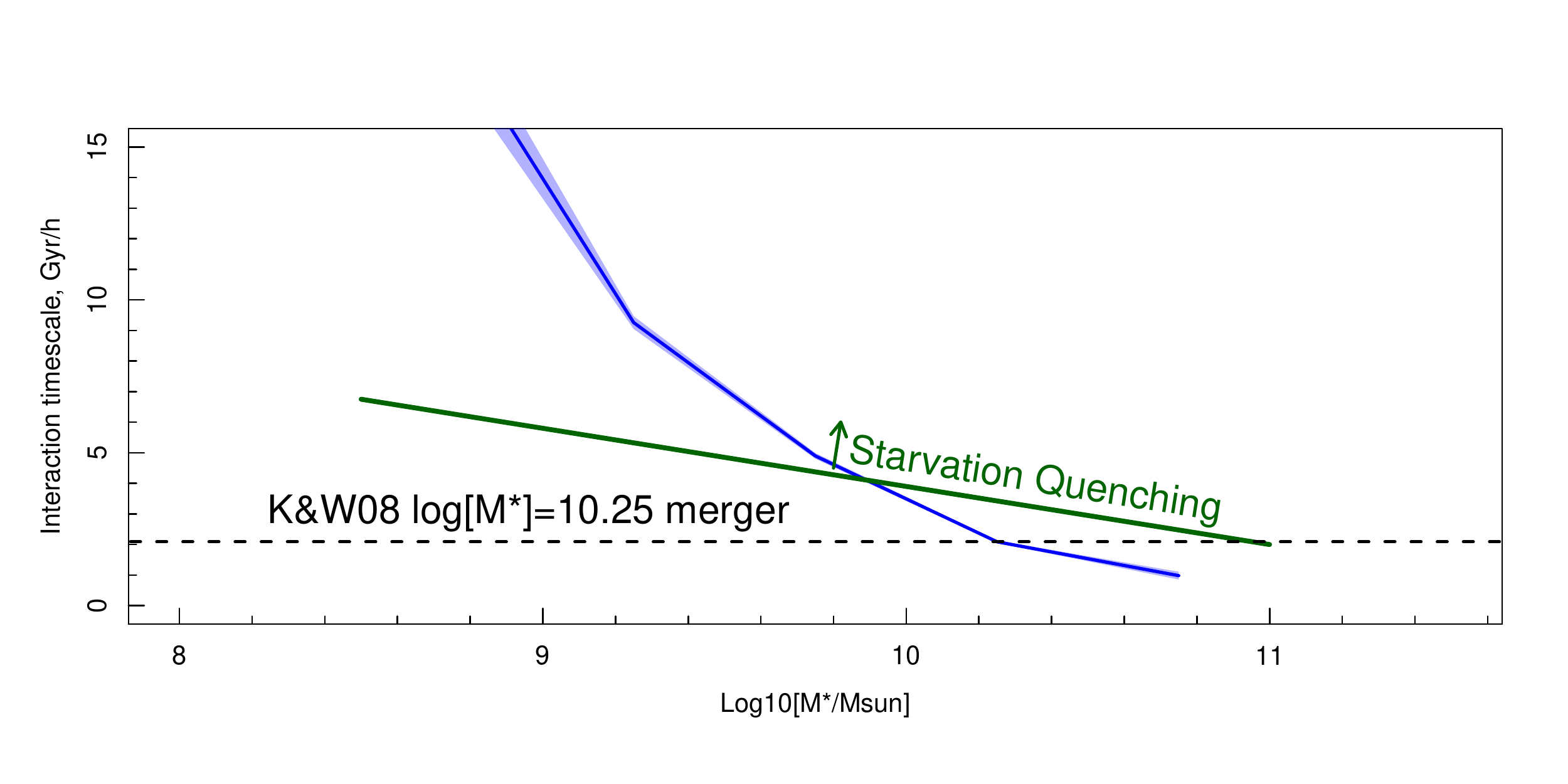}
\caption{Top: The fraction of satellite (secondary) pair galaxies which are identified as being passive as a function of stellar mass, scaled by the global fraction of all galaxies defined as passive. This gives a measure of how strongly interactions affect the production of passive galaxies at different stellar masses. At lower stellar masses we see a strong effect in interactions quenching star-formation in galaxies. Middle: the same as the top panel, but scaled by the passive fraction in non-pair galaxies. Bottom: the normalised, median interaction timescale, $\tau_{merge}$ as a function of stellar mass (see Section 5.2 for details). The coloured polygon displays the standard error on the median in each bin. We highlight the lower limit to the starvation quenching timescale as a function of stellar mass from \citet{Fillingham15} and the normalisation timescale of $\sim2$\,Gyr at log$_{10}$[M$_{*}$/M$_{\odot}$]=10.25 from \citet{Kitzbichler08}. Interaction times at log$_{10}$[M$_{*}$/M$_{\odot}$]$\lesssim$9.75 are sufficient to allow galaxies to become quenched via starvation. }
\label{fig:summary}
\end{center}
\end{figure*}

We calculate $\tau_{merge}$ for all satellite pair galaxies in our sample and take the median $\tau_{merge}$ in the same stellar mass bins as all other figures in this work. The bottom panel of Figure \ref{fig:summary} shows median $\tau_{merge}$ as a function of stellar mass normalised to the \cite{Kitzbichler08} mean interaction timescale at log$_{10}$[M$_{*}$/M$_{\odot}$]=10.25 (2.09\,Gyr). The coloured polygon displays the standard error on the median in each bin. The distribution of dynamical timescales and passive fraction in pairs are somewhat similar, with low-mass systems primarily having large merger timescales, the shortest timescales occurring at log$_{10}$[M$_{*}$/M$_{\odot}$]$\sim10.5$ and rising at higher stellar masses. Once again this point occurs close to the transition mass discussed previously. As such, we find that the point where passive fractions are almost identical between the field and groups/pairs (Figure \ref{fig:all_fractions}), where centrals and satellites in mergers have equal number density (Figure \ref{fig:all_fractions_split}), where the effect of interactions in forming quiescent galaxies and the merger timescale is shortest (Figure \ref{fig:summary}), and where major merger rates are highest \citep{Robotham14} all occur at roughly log$_{10}$[M$_{*}$/M$_{\odot}$]$\sim10.0-10.5$. Clearly this stellar mass is an important transition point between different galaxy-galaxy interaction driven/suppressed star-formation processes.            

Assuming this correlation between merger timescale and environmental quenching to be correct, we can begin to speculate about the processes which are causing these galaxies to become passive. It appears that interaction timescale is likely to be a driving factor in $observing$ passive low-mass satellite galaxies. Potentially galaxies going through interactions fail to replenish their gas reservoirs in the locally over-dense environment. As such, they have a fixed star-formation lifetime during the interaction. Interactions at around log$_{10}$[M$_{*}$/M$_{\odot}$]$\sim10.25$ happen quickly, and therefore the galaxies merge before we see them in their passive state. At lower stellar masses interaction timescales increase, thus, an increasingly larger fraction of galaxies have interaction timescales which are longer then their quenching times scales - and the passive fraction increases.        

Recent work by Peng et al (2015), found that the dominant process in turning galaxies passive is strangulation (galaxies can no longer draw from a reservoir of gas, use up all of their internal gas and become passive), and that this process occurs on timescales of $\sim4$\,Gyrs,  while \cite{Wheeler14, Fillingham15} find that quenching timescales of $>8$\,Gyr are required for intermediate mass satellites. Simulations, such as those of Boylan-Kolchin et al (2007), find that merger timescales can be as long as $>8$\,Gyr for low-mass minor mergers, and as short as $<$1\,Gyr for major mergers close to log$_{10}$[M$_{*}$/M$_{\odot}$]$\sim10.5$. Therefore, only low-mass satellites have the necessary interaction timescales to consume their gas, starve and reach their passive state. In the bottom panel of Figure \ref{fig:summary} we also over-plot the lower limit of the predicted starvation quenching timescale as a function of stellar mass, taken from Figure 6 of \cite{Fillingham15}. Using this simple comparison, we find that only galaxies at low (log$_{10}$[M$_{*}$/M$_{\odot}$]$<$10) stellar masses have interaction timescales sufficient for galaxies to become passive via starvation. At higher masses, timescales are short enough for galaxies to continue star-forming for the duration of the interaction, and not run out of star-forming gas. This is consistent with the comparative lack of significant gas consumption in mergers at these masses found by \citet{Ellison15} - post merger products primarily at $10<$log$_{10}$[M$_{*}$/M$_{\odot}$]$<$11 are found to have comparable HI gas fractions to non-merger control galaxies.

\begin{figure}
\begin{center}
\includegraphics[scale=0.30]{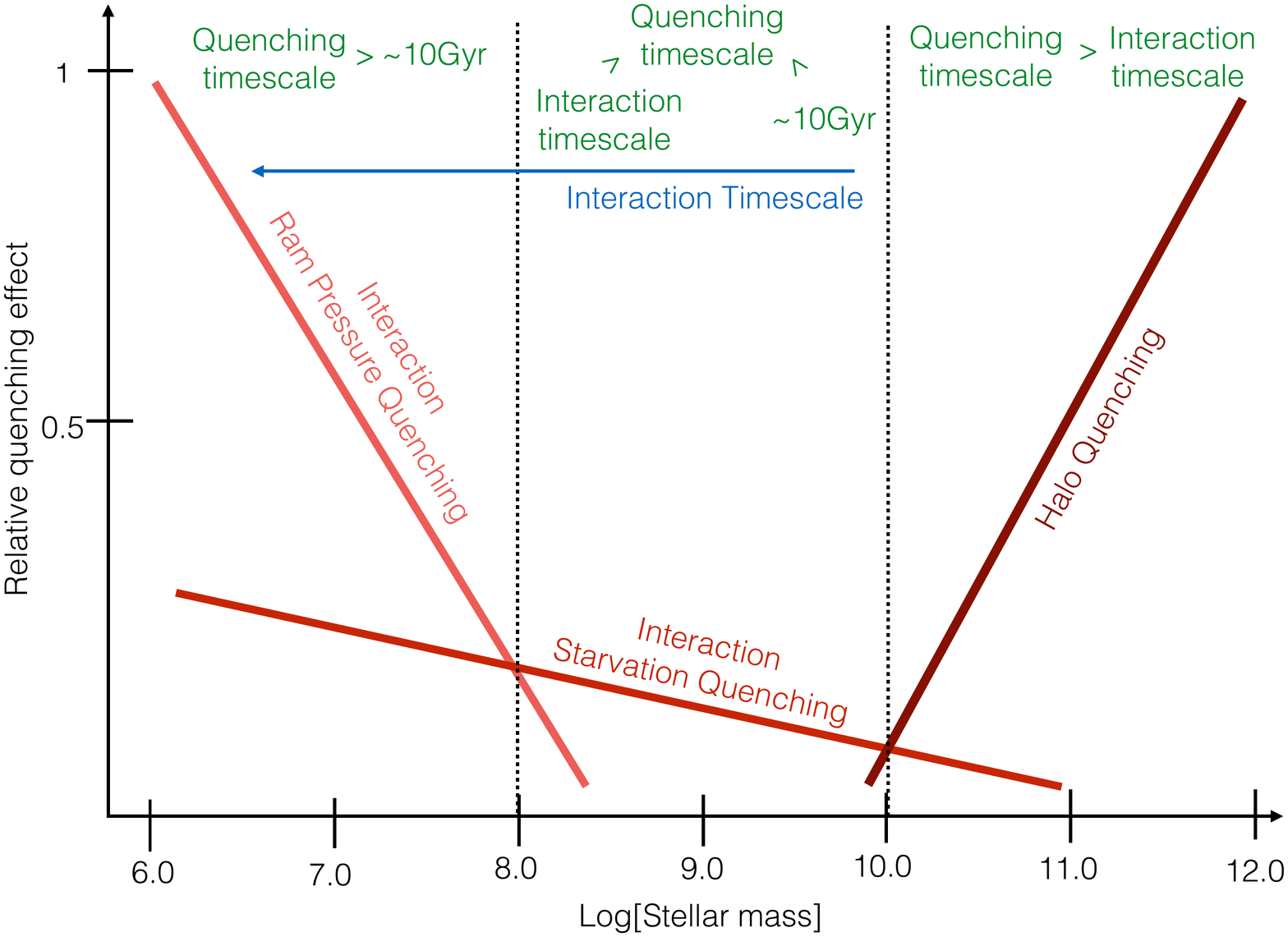}
\caption{Cartoon representation of the quenching processes affecting galaxies as a function of stellar mass. At low stellar masses galaxies are likely quenched via ram pressure stripping, at intermediate masses galaxies are sufficiently large not to be stripped, but become passive via strangulation in interactions (and this process becomes more dominant with decreasing stellar mass) and at high masses galaxies become passive via mass quenching, which is largely independent of galaxy-galaxy interactions. This figure is comparable to the passive fractions of galaxies outlined in Figure 5 of \citet{Weisz15}, combined with our results.  }
\label{fig:cartoon2}
\end{center}
\end{figure}

This simplistic picture, where interaction timescale governs the formation of passive low-mass galaxies, may be sufficient to explain the observed trends of passive fractions as a function of stellar mass in the $8<$log$_{10}$[M$_{*}$/M$_{\odot}$]$<$10 regime. Combining this with the previous work at both higher and lower stellar masses \citep[$e.g.$ see Figure 5 and 6 of][]{Weisz15}, we can build a picture of the relative contributions of different quenching effects as a function of stellar mass. A representation of this is shown in Figure \ref{fig:cartoon2}. In Local Group satellites at log$_{10}$[M$_{*}$/M$_{\odot}$]$<$8 \cite{Weisz15} see a sharp increase in the passive fraction with decreasing stellar mass. At these masses both starvation quenching timescales and interaction timescales are sufficiently long, that satellite galaxies have not merged but have also not had time to become passive via starvation. As such, the high passive fractions are attributed to efficient ram pressure stripping in low-mass galaxies, with the quenching efficiency increasing with decreasing stellar mass \citep[$i.e.$][]{Wheeler14,Wetzel15}. At 8$<$log$_{10}$[M$_{*}$/M$_{\odot}$]$<$10 (the masses of interest in this work) we see a low passive fraction, but a steadily increasing trend of the effect of interactions in turning galaxies passive with decreasing stellar mass. At these masses starvation quenching timescales are shorter than interaction timescales and satellites stay within an interaction long enough to become passive. At  log$_{10}$[M$_{*}$/M$_{\odot}$]$>$10 interactions are too short for galaxies to become passive via starvation quenching, however, passive fractions rise once again and increase to higher stellar masses. It is therefore likely that this quenching is not due to galaxy-galaxy interactions but driven by mass quenching processes.

\subsection{Simplistic model of star-formation in satellite galaxies at log$_{10}$[M$_{*}$/M$_{\odot}$]$>$8 }
\label{sec:toy}

To test this hypothesis further, we develop a simplistic model for the evolution of sSFRs in log$_{10}$[M$_{*}$/M$_{\odot}$]$>$8 satellites during interactions with various pair mass ratios. From \cite{Davies15} we know that interactions can cause both suppression and enhancement of star-formation in satellites, and in the previous section we discuss that this dichotomy may simply be a consequence of interaction timescale. As we wish to develop a model which fits all satellites, we require such a model to both enhance and suppress star-formation. 
 
\begin{figure*}
\begin{center}
\includegraphics[scale=0.8]{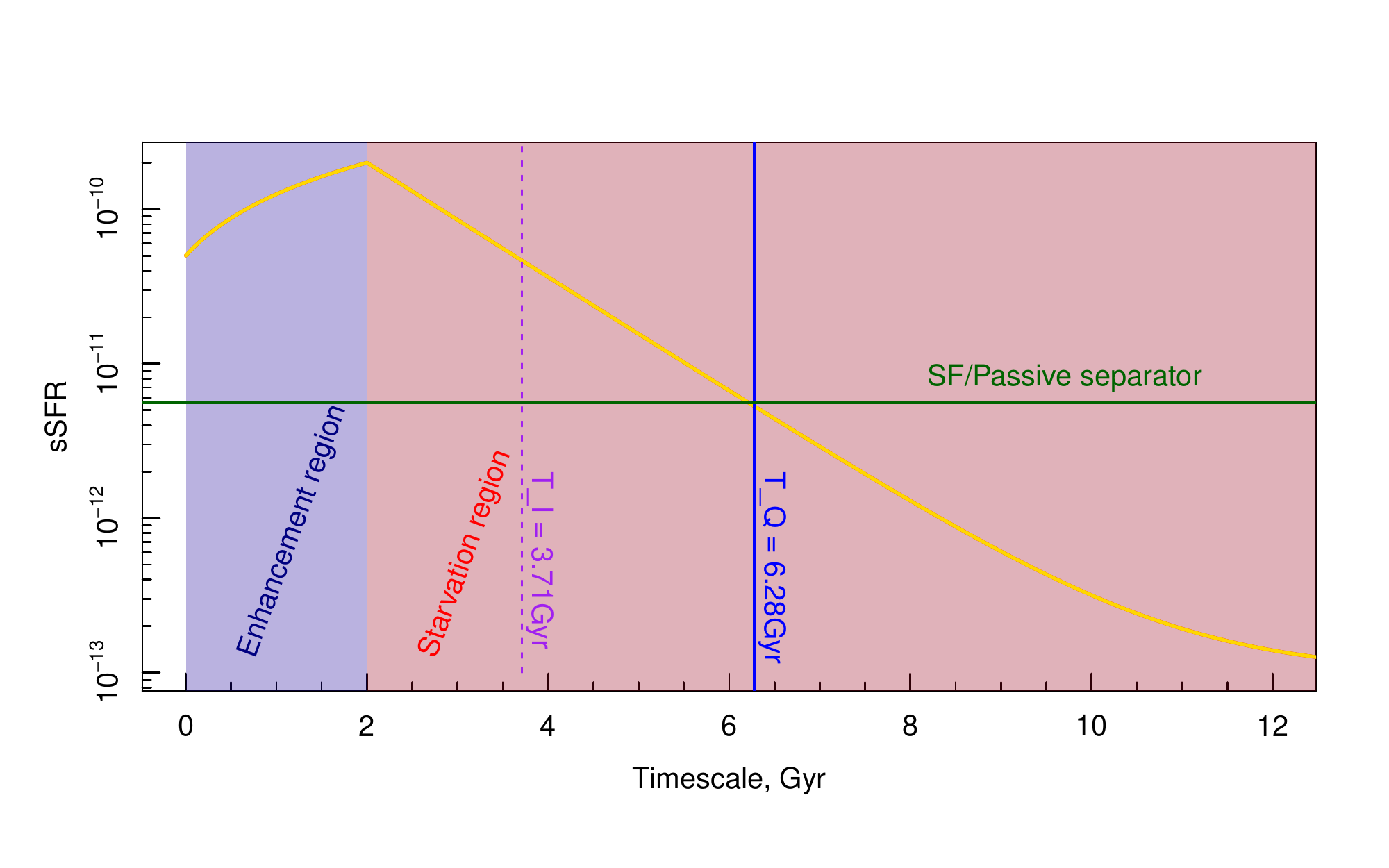}
\caption{A representation of the simple model used in this work to simulate satellite galaxies undergoing an interaction. The galaxy receives an initial enhancement to star-formation at the start of the interaction process and then a slow exponential decline (used to simulate starvation processes). The rate of this exponential decline is designed to ensure that galaxies are defined as passive when the exponential decline timescale reaches the predicted starvation quenching timescales of \citet{Fillingham15}. We display our SF/passive separator used in Section 3 to identify passive systems in the GAMA data. Our model is fixed so that it's sSFR crosses the separator line at the predicted starvation quenching timescale from \citet{Fillingham15}, T$_{Q}$. We then also display  the merger timescale for each modelled galaxy, T$_{merge}$, as the vertical dashed purple line.}
\label{fig:model}
\end{center}
\end{figure*}

We initially start with the premise that the sSFR is enhanced at the early stages of the interaction, and then exponentially declines after some time (simulating the slow strangulation scenario). We set the sSFR of a galaxy upon entering an interaction as the median sSFR of all non-pair galaxies in our data at $\pm0.05$dex in stellar mass from our model galaxy. In \cite{Davies15} we find that on average major mergers at log$_{10}$[M$_{*}$/M$_{\odot}$]=10-10.5 show $\sim$4 times increase in their sSFR. Using the merger timescale estimates of \cite{Kitzbichler08}, we predict these interactions to last $\sim2$\,Gyr. We treat these mergers as us witnessing the initial stages of all mergers, and linearly increase the sSFR of our model galaxy by 4-5 times (randomly selected) over the first 2\,Gyr of the interaction. We then assume that at $>2$\,Gyr the sSFR exponentially declines as the galaxy become starved. In order to predict the rate of decline we use the minimum quenching timescales from \cite{Fillingham15} and set our exponential decay so that the galaxy becomes passive using our SF/passive selection (log$_{10}$[sSFR]$<$-11.25) at the quenching timescale corresponding to its stellar mass ($i.e.$ for a given stellar mass satellite, our model is defined as passive at the quenching time defined by Fillingham et al). This naturally produces different exponential decay slopes as a function of stellar mass. We derive an analytic form for sSFR in an interaction which meets the above criteria as follows:

\begin{scriptsize}
\begin{equation}
sSFR[\tau<=2Gyr]=(\frac{(Q-1)\tau}{2}+1)(sSFR_{0}-sSFR_{Floor})+sSFR_{Floor}, \\
\end{equation}
 \begin{equation}
sSFR[\tau>2Gyr]= \frac{Q\tau_{char}}{e[\frac{3.3S_{char}\tau}{\tau_{Q}}]} (sSFR_{0}-sSFR_{Floor})+sSFR_{Floor}
\end{equation}
\end{scriptsize}

\noindent where, $Q$ = initial sSFR enchantment factor \citep[4-5 from][]{Davies15}, $\tau$ = timescale within the interaction, sSFR$_{0}$ = starting sSFR defined from all non-pair galaxies at $\pm0.1$dex in stellar mass, sSFR$_{Floor}$ = sSFR at $\tau$ = infinity which we set to log[sSFR$_{Floor}$]=-13.0, $\tau_{char}$ = characteristic decay rate ($e^{[6.6S_{char}/\tau_{Q}]}$), $\tau_{Q}$ = quenching timescale from \cite{Fillingham15}, which has the form $\tau_{Q}$= -1.9log$_{10}$[M$_{*}$/M$_{\odot}$]+22.9, and $S_{char}$ = characteristic sSFR ($sSFR_{char}=[sSFR_{0}/10^{-10.5}]^{0.22}$), which is used to match the starvation quenching point from \cite{Fillingham15} to our SF/passive separation point. Figure \ref{fig:model} shows sSFR against interaction timescale from our model for a  log$_{10}$[M$_{*}$/M$_{\odot}$]$\sim9.8$ galaxy interacting with a log$_{10}$[M$_{*}$/M$_{\odot}$]$\sim10.8$ galaxy with log$_{10}$[sSFR$_{0}$]=-10.3. We highlight the initial star-formation  enhancement region and subsequent starvation region. We also over-plot the quenching point (T$_{Q}$) defined by the minimum quenching timescale from \cite{Fillingham15} after the 2\,Gyr point and the SF/passive separator line used in this work.

Following  this, we wish to define a timescale over which this system could be observed as an interacting pair galaxy. To do this we use the merger timescale ($\tau_{merge}$) predictions used in the previous section, and set an upper limit of 13\,Gyr (systems can not have been in an interaction for longer than this, given the age of the Universe). The vertical purple dashed line in Figure \ref{fig:model} displays $\tau_{merge}$ for this model, and as such, this system could only be observed as an interacting pair at times shorter than this point. Clearly, if this model system is observed as an interacting pair it will be classed as star-forming, as the quenching timescale is much longer than $\tau_{merge}$.

In order to test the validity of this model we use it to make a prediction for the sSFR in interacting systems, calculate the predicted passive fraction, and relate these to the observed passive fraction in GAMA. To do this we take a random point on our model at $\tau<\tau_{merge}$ and treat this as the interacting galaxy's sSFR. We take all true satellite pair galaxies in our sample, calculate sSFR$_{0}$ from the median sSFR of all non-pair galaxies at $\pm0.05$dex in stellar mass, and use their stellar mass and pair mass ratio to calculate a model sSFR vs interaction timescale. We then randomly select a sSFR at $\tau<\tau_{merge}$. This provides a predicted sSFR for all satellite galaxies in our data assuming our model to be correct. From this we calculate a passive fraction as a function of stellar mass, as in the previous section. We repeat this process 500 times to quantify any biases in the random selection at times below $\tau_{merge}$.  

\begin{figure*}
\begin{center}
\includegraphics[scale=0.65]{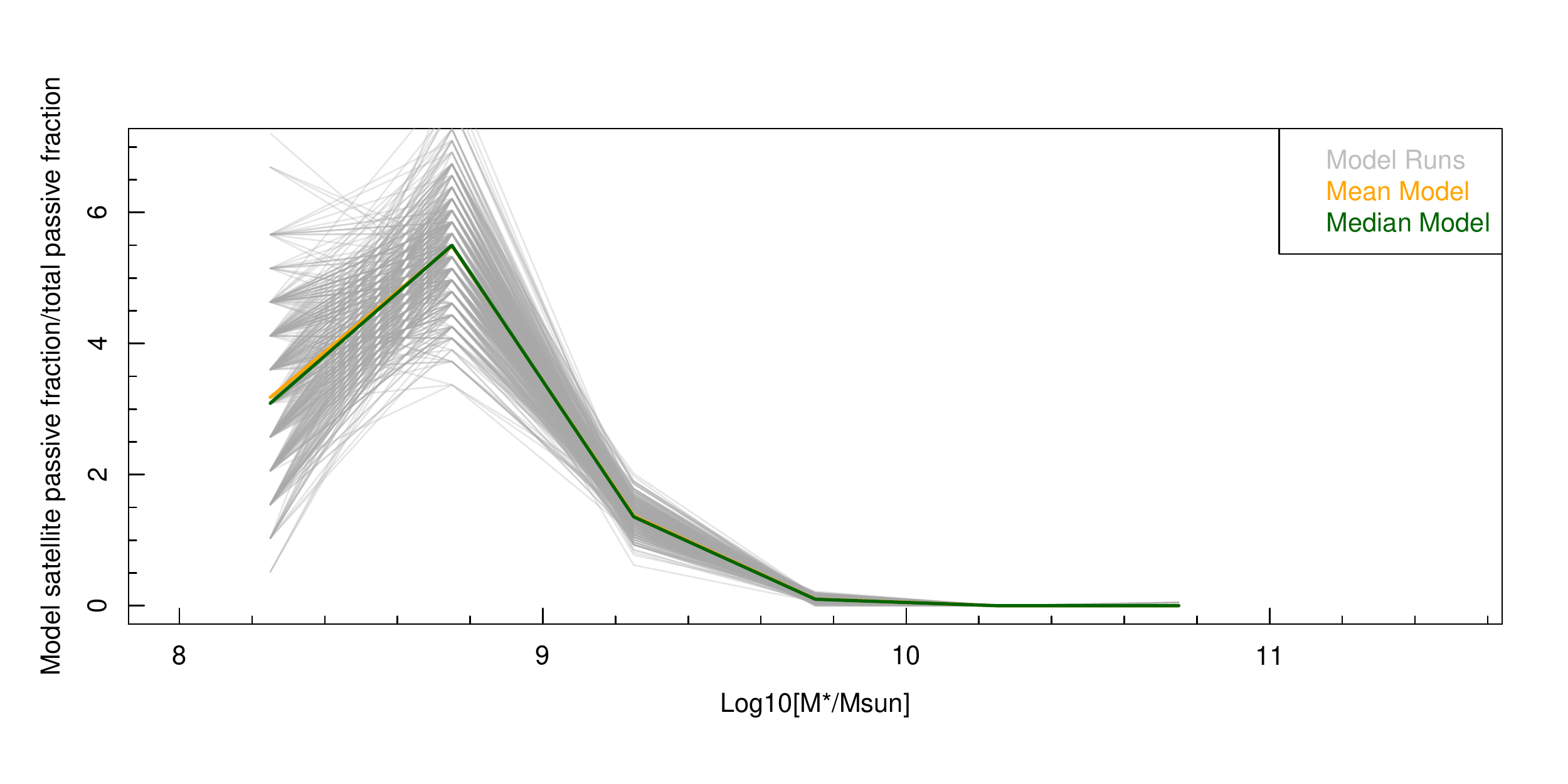}
\caption{Model run passive fractions as a function of stellar mass. We produce our simplistic model for all satellite galaxies in our sample, and predict their sSFR during the interaction (see Section 5.3 for details). We then calculate the model passive fraction as a function of stellar mass as in Figure \ref{fig:summary}. Our modelling process contains a random observation time, as such we repeat our model runs 500 times. The grey lines show the passive fraction from each of these runs - and therefore, the spread of these lines gives and estimate of the random error in the passive fraction from our toy model. The gold and green lines show the mean and median of these runs respectively.}
\label{fig:model_fraction}
\end{center}
\end{figure*}

Figure \ref{fig:model_fraction} displays the model passive fraction as a function of stellar mass for all runs, and the mean/median of the runs. Comparing this to the top panel of Figure \ref{fig:summary} (the true distribution in the data), we find that our model predictions show similar characteristics, they are both low at log$_{10}$[M$_{*}$/M$_{\odot}$]=10.5 and rise with deceasing stellar mass to  log$_{10}$[M$_{*}$/M$_{\odot}$]$\sim$8.75, and then slightly drop to log$_{10}$[M$_{*}$/M$_{\odot}$]$\sim$8.25. Our model passive fractions are higher at log$_{10}$[M$_{*}$/M$_{\odot}$]$<$9.0 and lower at log$_{10}$[M$_{*}$/M$_{\odot}$]$>$9.0, but are largely similar in shape to the observed distribution. The drop in passive fractions at the lowest stellar masses is interesting, and is seen in both the data and models. This drop is potentially due to quenching timescales approaching a considerable fraction of the age of the Universe. At low stellar masses only satellites which began an interaction in the very early Universe will have sufficient time to become quenched via starvation. This is in fact one of the reasons for proposing the ram pressure quenching scenario at the lowest stellar masses - low-mass Local Group galaxies do not have sufficient time to become passive by any other method.  

While this model is simplistic and contains speculative assumptions, it is largely consistent with the observed passive fractions. It provides a model where interaction timescale is the governing factor in the observed passive fractions as a function of stellar mass, and roughly produces the necessary distribution of passive fractions in satellite galaxies  - being low at  log$_{10}$[M$_{*}$/M$_{\odot}$]$\sim$10 and consistently rising to lower stellar masses, and then plateauing/dropping at log$_{10}$[M$_{*}$/M$_{\odot}$]$<$8.5. 

\begin{figure*}
\begin{center}
\includegraphics[scale=0.65]{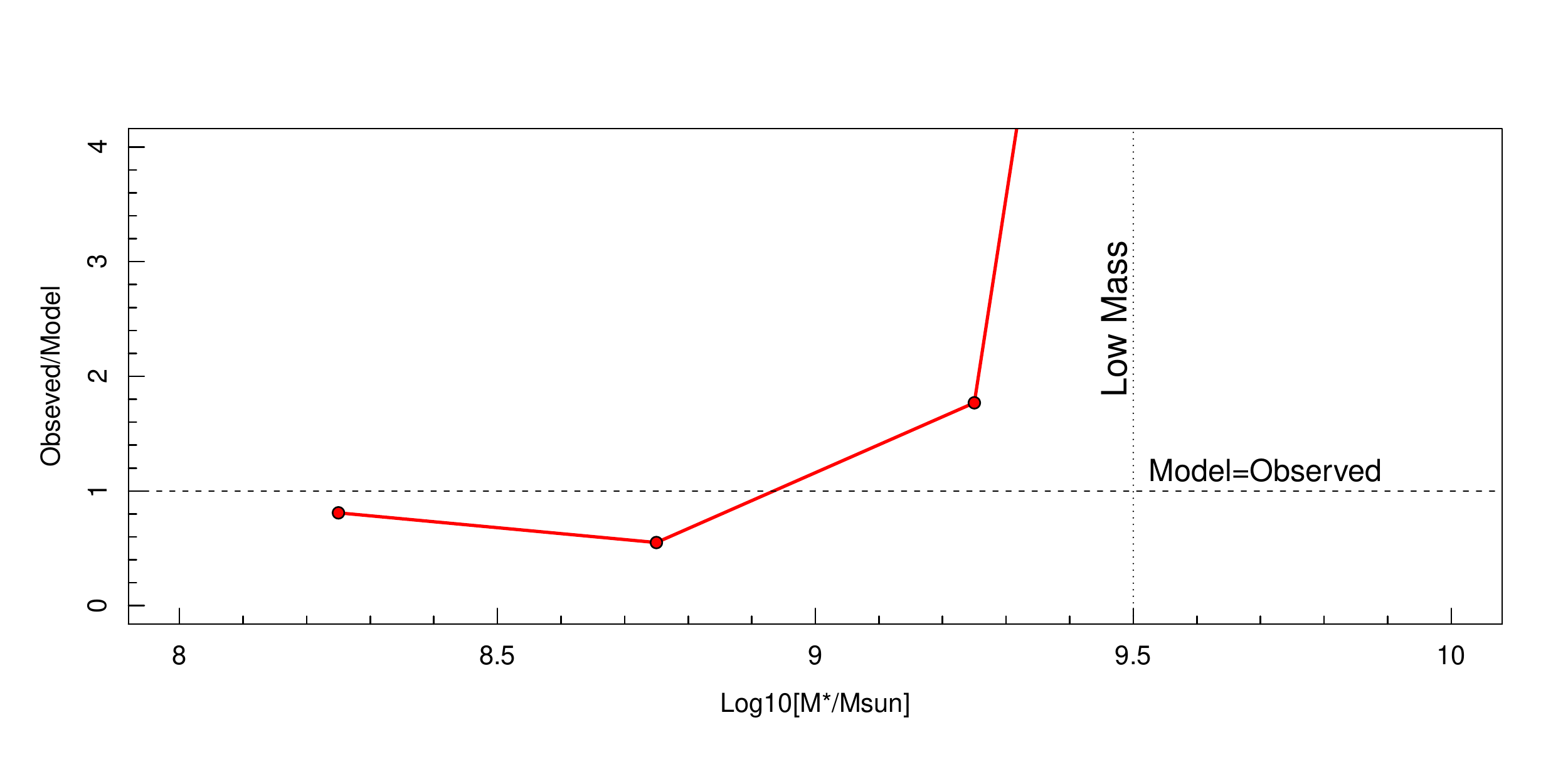}
\caption{Comparison between the passive fraction in the observed GAMA data (Figure \ref{fig:summary}) and that produced from our toy model (Figure \ref{fig:model_fraction}) in the low mass regime. The line shows the observed GAMA distribution divided by the mean distribution from our toy model runs. At log$_{10}$[M$_{*}$/M$_{\odot}$]$<$9.5 our toy model is consistent with the observed distribution (the mass regime where our starvation model should be appropriate).}
\label{fig:model_fraction_vs_obs}
\end{center}
\end{figure*}

To compare our model passive fractions with the observed passive fractions, Figure \ref{fig:model_fraction_vs_obs} shows the ratio of the two distributions, in the low stellar mass regime. The solid line and points display the observed passive fraction divided by the mean model passive fraction (gold line in Figure \ref{fig:model_fraction}). At 8.0$<$log$_{10}$[M$_{*}$/M$_{\odot}$]$<$9.5 the model is roughly consistent with the observed distribution (within a factor of two in the region of interest in this work). At higher stellar masses the model passive fraction is essentially zero, and can not be directly compared to the observed distribution - our model only probes starvation quenching scenarios, which are likely to only be appropriate in the 8.0$<$log$_{10}$[M$_{*}$/M$_{\odot}$]$<$10 regime.

\section{Conclusions \& Future Prospects}       

We have investigated the environmental distribution of passive and star-forming galaxies in the GAMA sample. We find that passive fractions are higher in group/pair environments than isolated environments for all stellar masses, suggesting that group/pair interactions have a significant role in producing quiescent systems.  We find that the global passive fractions drop with decreasing stellar mass in all environments until log$_{10}$[M$_{*}$/M$_{\odot}$]$=9$. For pair galaxies we see an increase in the passive fraction in our lowest stellar mass bin, potentially indicating that we are witnessing the increase in passive fraction predicted from ram pressure quenching in low-mass galaxies \citep{Fillingham15}. We highlight that log$_{10}$[M$_{*}$/M$_{\odot}$]$\sim10.25$ is an important transition mass, below which interactions have a significant role in suppressing star-formation in galaxies.

We investigate the passive fraction of pair galaxies in comparison to the global passive fraction, and show that the pair passive fraction increases with decreasing stellar mass, suggesting the effect of interactions in producing passive galaxies becomes more significant at lower stellar masses. However, we highlight that this trend is similar to the changes in interaction timescale with stellar mass. We propose, consistently with other authors, that the formation of passive galaxies at log$_{10}$[M$_{*}$/M$_{\odot}$]$<10.25$ primarily occurs via starvation due to an interaction. The interaction initially boosts star formation in the satellite galaxy, but inhibits the system from replenishing its gas (or potentially such gas in the local environment has already been accreted by its larger companion). As such, the galaxy consumes all of its gas, starves and become passive over an extended time period. 

We suggest that the increased passive fraction at low stellar masses are simply due to extended interaction timescales. Essentially, all galaxies would eventually become passive in an interaction, however this process will occur over long timescales ($>8$\,Gyr). Interactions of galaxies around log$_{10}$[M$_{*}$/M$_{\odot}$]$=10$ are exceptionally short ($<$1\,Gyr) and therefore, the galaxies merge prior to their passive phase. For less massive galaxies, interaction timescales are dramatically increased, and a large fraction of galaxies reach their passive phase prior to merging. This causes us to observe an increased passive fraction with decreasing stellar mass. 

We produce a simplistic model, where satellites have an initial, short duration, burst of star-formation and then a exponential decline in star-formation (used to mimic the starvation processes). This model largely reproduces the shape of the observed satellite passive fraction.  

The prospects for future studies to probe this scenario are tantalising. Combinations of the COSMOS field data \citep{Scoville07} and the upcoming COSMOS HI Large Extragalactic Survey \citep[CHILES see,][]{Fernandez13}, the Looking At the Distant Universe with the MeerKat Array (LADUMA, Holwerda in prep) with deep spectroscopic surveys in the Chandra Deep Filed South (CDFS) region, and GAMA combined with the Australia Square Kilometre Array Pathfinder (ASKAP) Deep Investigation of Neutral Gas Origins Survey \cite[DINGOS,][]{Meyer09}, will give us key insights into the HI content of low-mass mergers - potentially allowing us to directly probe gas depletion in these galaxies. It will also be interesting to consider the varying gas-to-stellar mass ratio at low stellar masses \cite[$e.g.$][]{Kannappan09}, and how this ratio changes with local environment. 

Looking further ahead the Wide Area VISTA Extragalactic Survey \citep[WAVES,][]{Driver15} will obtain spectra for $\sim1$\,Million galaxies at $z<0.2$ probing stellar mass limits down to log$_{10}$[M$_{*}$/M$_{\odot}$]$>6.5$, and fully bridging the transition between previous samples of local low-mass galaxies and Local Group systems - fully mapping the potential transition between starvation quenched and ram-pressure quenched systems.

\section*{Acknowledgements}

GAMA is a joint European-Australasian project based around a spectroscopic campaign using the Anglo-Australian Telescope. The GAMA input catalogue is based on data taken from the Sloan Digital Sky Survey and the UKIRT Infrared Deep Sky Survey. Complementary imaging of the GAMA regions is being obtained by a number of independent survey programs including GALEX MIS, VST KiDS, VISTA VIKING, WISE, Herschel-ATLAS, GMRT and ASKAP providing UV to radio coverage. GAMA is funded by the STFC (UK), the ARC (Australia), the AAO, and the participating institutions. The GAMA website is http://www.gama-survey.org/. SB acknowledges the funding support from the Australian Research Council through a Future Fellowship (FT140101166). M.S.O acknowledges the funding support of the Australian Research Council through a Future Fellowship (FT140100255). Mahajan is a Fast track fellow at IISER, Mohali funded by the Department of Science (DST) Science Education and Research Board (SERB) grant number SB/FTP/PS-054/2013. We thank the anonymous referee for their extremely helpful comments and suggestions in improving the overall quality and clarity of this paper.


\begin{thebibliography}{}

\bibitem[Ahn et al.(2014)]{Ahn14} Ahn, C.~P., Alexandroff, 
R., Allende Prieto, C., et al.\ 2014, \apjs, 211, 17 

\bibitem[\protect\citeauthoryear{Alpaslan et 
al.}{2015}]{Alpaslan15} Alpaslan M., et al., 2015, MNRAS, 451, 
3249


\bibitem[\protect\citeauthoryear{Baldry et al.}{2006}]{Baldry06} 
Baldry I.~K., Balogh M.~L., Bower R.~G., Glazebrook K., Nichol R.~C., 
Bamford S.~P., Budavari T., 2006, MNRAS, 373, 469 


\bibitem[Baldry et al.(2010)]{Baldry10} Baldry, I.~K., Robotham, 
A.~S.~G., Hill, D.~T., et al.\ 2010, \mnras, 404, 86

\bibitem[\protect\citeauthoryear{Balogh et al.}{2004}]{Balogh04} 
Balogh M.~L., Baldry I.~K., Nichol R., Miller C., Bower R., Glazebrook K., 
2004, ApJ, 615, L101 

\bibitem[\protect\citeauthoryear{Behroozi et 
al.}{2015}]{Behroozi15} Behroozi P.~S., et al., 2015, MNRAS, 450, 
1546 


\bibitem[Baldwin et al.(1981)]{Bladwin81} Baldwin, J.~A., 
Phillips, M.~M., \& Terlevich, R.\ 1981, PASA, 93, 5 

\bibitem[\protect\citeauthoryear{Boylan-Kolchin 
\& Ma}{2007}]{Boylan-Kolchin07} Boylan-Kolchin M., Ma C.-P., 2007, MNRAS, 374, 1227 

\bibitem[\protect\citeauthoryear{Brough et al.}{2013}]{Brough13} 
Brough S., et al., 2013, MNRAS, 435, 2903 


\bibitem[\protect\citeauthoryear{Cameron}{2011}]{Cameron11} 
Cameron E., 2011, PASA, 28, 128 

\bibitem[\protect\citeauthoryear{Chabrier}{2003}]{Chabrier03} 
Chabrier G., 2003, PASP, 115, 763 

\bibitem[\protect\citeauthoryear{Cowie et al.}{1996}]{Cowie96} 
Cowie L.~L., Songaila A., Hu E.~M., Cohen J.~G., 1996, AJ, 112, 839 

\bibitem[da Cunha et al.(2008)]{daCunha08} da Cunha, E., Charlot, 
S., \& Elbaz, D.\ 2008, \mnras, 388, 1595 

\bibitem[Dav{\'e}(2008)]{Dave08} Dav{\'e}, R.\ 2008, \mnras, 
385, 147 

\bibitem[\protect\citeauthoryear{Davies 
\& Phillips}{1989}]{Davies88} Davies J.~I., Phillips S., 1989, Ap\&SS, 157, 291 

\bibitem[Davies et al.(2015)]{Davies15} Davies, L. ~J. ~M, et al.\ 2015, \mnras, X, X

\bibitem[\protect\citeauthoryear{Drinkwater et 
al.}{2001}]{Drinkwater01} Drinkwater M.~J., Gregg M.~D., Holman 
B.~A., Brown M.~J.~I., 2001, MNRAS, 326, 1076 

\bibitem[Driver et al.(2011)]{Driver11} Driver, S.~P., Hill, 
D.~T., Kelvin, L.~S., et al.\ 2011, \mnras, 413, 971 


\bibitem[Driver et al.(2013)]{Driver13} Driver, S.~P., Robotham, 
A.~S.~G., Bland-Hawthorn, J., et al.\ 2013, \mnras, 430, 2622 



\bibitem[Driver et al. (2015)]{Driver15}, Driver S. et al.,  In ?The Universe of Digital Sky
Surveys?, Naples, Italy, Nov. 2014


\bibitem[\protect\citeauthoryear{Driver et al.}{2015}]{Driver15b} 
Driver S.~P., et al., 2015, arXiv, arXiv:1508.02076 



\bibitem[\protect\citeauthoryear{Ellison et 
al.}{2008}]{Ellison08} Ellison S.~L., Patton D.~R., Simard L., 
McConnachie A.~W., 2008, AJ, 135, 1877 

\bibitem[\protect\citeauthoryear{Ellison et 
al.}{2010}]{Ellison10} Ellison S.~L., Patton D.~R., Simard L., 
McConnachie A.~W., Baldry I.~K., Mendel J.~T., 2010, MNRAS, 407, 1514 

\bibitem[\protect\citeauthoryear{Ellison et 
al.}{2015}]{Ellison15} Ellison S.~L., Fertig D., Rosenberg J.~L., 
Nair P., Simard L., Torrey P., Patton D.~R., 2015, MNRAS, 448, 221 


\bibitem[Fern{\'a}ndez et al.(2013)]{Fernandez13} Fern{\'a}ndez, 
X., van Gorkom, J.~H., Hess, K.~M., et al.\ 2013, arXiv:1303.2659 

\bibitem[\protect\citeauthoryear{Fillingham et 
al.}{2015}]{Fillingham15} Fillingham S.~P., Cooper M.~C., Wheeler 
C., Garrison-Kimmel S., Boylan-Kolchin M., Bullock J.~S., 2015, arXiv, 
arXiv:1503.06803


\bibitem[\protect\citeauthoryear{Gao 
\& White}{2007}]{Gao07} Gao L., White S.~D.~M., 2007, MNRAS, 377, L5 

\bibitem[\protect\citeauthoryear{Geha et al.}{2012}]{Geha12} 
Geha M., Blanton M.~R., Yan R., Tinker J.~L., 2012, ApJ, 757, 85 

\bibitem[\protect\citeauthoryear{Grebel, Gallagher, 
\& Harbeck}{2003}]{Grebel03} Grebel E.~K., Gallagher J.~S., III, Harbeck D., 2003, AJ, 125, 1926 


\bibitem[Gunawardhana et al.(2011)]{Gunawardhana11} Gunawardhana, 
M.~L.~P., Hopkins, A.~M., Sharp, R.~G., et al.\ 2011, \mnras, 415, 1647 


\bibitem[\protect\citeauthoryear{Haines et al.}{2007}]{Haines07} 
Haines C.~P., Gargiulo A., La Barbera F., Mercurio A., Merluzzi P., 
Busarello G., 2007, MNRAS, 381, 7 

\bibitem[\protect\citeauthoryear{Haines, Gargiulo, \& Merluzzi}{2008}]{Haines08} Haines C.~P., Gargiulo A., Merluzzi P., 2008, MNRAS, 385, 1201 



\bibitem[\protect\citeauthoryear{Hearin, Watson, 
\& van den Bosch}{2015}]{Hearin15} Hearin A.~P., Watson D.~F., van den Bosch F.~C., 2015, MNRAS, 452, 1958


\bibitem[\protect\citeauthoryear{Hearin \& Watson}{2013}]{Hearin13} Hearin A.~P., Watson D.~F., 2013, MNRAS, 435, 1313



\bibitem[\protect\citeauthoryear{Hopkins et 
al.}{2013}]{Hopkins13} Hopkins A.~M., et al., 2013, MNRAS, 430, 
2047 

\bibitem[Hopkins et al.(2013)]{Hopkins13} Hopkins, A.~M., Driver, 
S.~P., Brough, S., et al.\ 2013, \mnras, 430, 2047 



\bibitem[\protect\citeauthoryear{Kannappan, Guie, 
\& Baker}{2009}]{Kannappan09} Kannappan S.~J., Guie J.~M., Baker A.~J., 2009, AJ, 138, 579 

\bibitem[\protect\citeauthoryear{Kauffmann et 
al.}{2013}]{Kauffmann13} Kauffmann G., Li C., Zhang W., Weinmann 
S., 2013, MNRAS, 430, 1447 


\bibitem[Kennicutt(1998)]{Kennicutt98} Kennicutt, R.~C., Jr.\ 1998, 
\apj, 498, 541 


\bibitem[\protect\citeauthoryear{Kimm et al.}{2009}]{Kimm09} 
Kimm T., et al., 2009, MNRAS, 394, 1131

\bibitem[\protect\citeauthoryear{Kitzbichler \& White}{2008}]{Kitzbichler08} Kitzbichler M.~G., White S.~D.~M., 2008, MNRAS, 391, 1489

\bibitem[\protect\citeauthoryear{Kroupa}{2001}]{Kroupa01} Kroupa 
P., 2001, MNRAS, 322, 231 

\bibitem[\protect\citeauthoryear{Lange et al.}{2015}]{Lange15} 
Lange R., et al., 2015, MNRAS, 447, 2603 

\bibitem[\protect\citeauthoryear{Lara-L{\'o}pez et 
al.}{2013}]{Lara-lopez13} Lara-L{\'o}pez M.~A., et al., 2013, MNRAS, 
434, 451


\bibitem[Lewis et al.(2002)]{Lewis02} Lewis, I.~J., Cannon, 
R.~D., Taylor, K., et al.\ 2002, \mnras, 333, 279 

\bibitem[\protect\citeauthoryear{Liske et al.}{2015}]{Liske15} 
Liske J., et al., 2015, arXiv, arXiv:1506.08222 

\bibitem[\protect\citeauthoryear{Mahajan, Haines, \& Raychaudhury}{2010}]{Mahajan10} Mahajan S., Haines C.~P., Raychaudhury S., 2010, MNRAS, 404, 1745 

\bibitem[\protect\citeauthoryear{Mahajan et al.}{2015}]{Mahajan15} Mahajan S., et al., 2015, MNRAS, 446, 2967


\bibitem[\protect\citeauthoryear{Martig et al.}{2009}]{Martig09} 
Martig M., Bournaud F., Teyssier R., Dekel A., 2009, ApJ, 707, 250 

\bibitem[\protect\citeauthoryear{Martin et al.}{2007}]{Martin07} 
Martin D.~C., et al., 2007, ApJS, 173, 342 

\bibitem[\protect\citeauthoryear{Meyer}{2009}]{Meyer09} Meyer 
M., 2009, pra..conf, 15 

\bibitem[\protect\citeauthoryear{Patton et al.}{2011}]{Patton11} 
Patton D.~R., Ellison S.~L., Simard L., McConnachie A.~W., Mendel J.~T., 
2011, MNRAS, 412, 591 


\bibitem[\protect\citeauthoryear{Peng et al.}{2010}]{Peng10} Peng Y.-j., et al., 2010, ApJ, 721, 193


\bibitem[\protect\citeauthoryear{Peng et al.}{2012}]{Peng12} 
Peng Y.-j., Lilly S.~J., Renzini A., Carollo M., 2012, ApJ, 757, 4 

\bibitem[\protect\citeauthoryear{Peng, Maiolino, \& Cochrane}{2015}]{Peng15} Peng Y., Maiolino R., Cochrane R., 2015, Natur, 521, 192 


\bibitem[\protect\citeauthoryear{Phillips et 
al.}{2014}]{Phillips14} Phillips J.~I., Wheeler C., Boylan-Kolchin 
M., Bullock J.~S., Cooper M.~C., Tollerud E.~J., 2014, MNRAS, 437, 1930 


\bibitem[\protect\citeauthoryear{Phillips et 
al.}{2015}]{Phillips15} Phillips J.~I., Wheeler C., Cooper M.~C., 
Boylan-Kolchin M., Bullock J.~S., Tollerud E., 2015, MNRAS, 447, 698 





\bibitem[Robotham et al.(2010)]{Robotham10} Robotham, A., Driver, S.~P., Norberg, P., et al.\ 2010, PASA, 27, 76 

\bibitem[Robotham et al.(2011)]{Robotham11} Robotham, A.~S.~G., 
Norberg, P., Driver, S.~P., et al.\ 2011, \mnras, 416, 2640

\bibitem[Robotham et al.(2012)]{Robotham12} Robotham, A.~S.~G., 
Baldry, I.~K., Bland-Hawthorn, J., et al.\ 2012, \mnras, 424, 1448 


\bibitem[Robotham et al.(2013)]{Robotham13} Robotham, A.~S.~G., 
Liske, J., Driver, S.~P., et al.\ 2013, \mnras, 431, 167 

\bibitem[\protect\citeauthoryear{Robotham et 
al.}{2014}]{Robotham14} Robotham A.~S.~G., et al., 2014, MNRAS, 
444, 3986 



 \bibitem[Saunders et al.(2004)]{Saunders04} Saunders, W., Bridges, 
T., Gillingham, P., et al.\ 2004, \procspie, 5492, 389 


 \bibitem[Scoville et al.(2007)]{Scoville07} Scoville, N., Aussel, 
H., Brusa, M., et al.\ 2007, \apjs, 172, 1 

\bibitem[\protect\citeauthoryear{Scudder et 
al.}{2012}]{Scudder12} Scudder J.~M., Ellison S.~L., Torrey P., 
Patton D.~R., Mendel J.~T., 2012, MNRAS, 426, 549
 
 
 \bibitem[Sharp et al.(2006)]{Sharp06} Sharp, R., Saunders, W., 
Smith, G., et al.\ 2006, \procspie, 6269, 62690G 
 

 
\bibitem[Taylor et al.(2011)]{Taylor11} Taylor, E.~N., Hopkins, 
A.~M., Baldry, I.~K., et al.\ 2011, \mnras, 418, 1587 
 
 \bibitem[\protect\citeauthoryear{Taylor et al.}{2015}]{Taylor15} 
Taylor E.~N., et al., 2015, MNRAS, 446, 2144 
 
 
\bibitem[\protect\citeauthoryear{Tinker et al.}{2013}]{Tinker13} 
Tinker J.~L., Leauthaud A., Bundy K., George M.~R., Behroozi P., Massey R., 
Rhodes J., Wechsler R.~H., 2013, ApJ, 778, 93 
 
 \bibitem[\protect\citeauthoryear{Tinsley}{1968}]{Tinsley68} 
Tinsley B.~M., 1968, ApJ, 151, 547 


\bibitem[\protect\citeauthoryear{Weisz et al.}{2015}]{Weisz15} 
Weisz D.~R., Dolphin A.~E., Skillman E.~D., Holtzman J., Gilbert K.~M., 
Dalcanton J.~J., Williams B.~F., 2015, ApJ, 804, 136

\bibitem[\protect\citeauthoryear{Wetzel, Tollerud, 
\& Weisz}{2015}]{Wetzel15} Wetzel A.~R., Tollerud E.~J., Weisz D.~R., 2015, ApJ, 808, L27 

\bibitem[\protect\citeauthoryear{Wetzel et al.}{2013}]{Wetzel13} 
Wetzel A.~R., Tinker J.~L., Conroy C., van den Bosch F.~C., 2013, MNRAS, 
432, 336


\bibitem[\protect\citeauthoryear{Wheeler et 
al.}{2014}]{Wheeler14} Wheeler C., Phillips J.~I., Cooper M.~C., 
Boylan-Kolchin M., Bullock J.~S., 2014, MNRAS, 442, 1396


 \bibitem[\protect\citeauthoryear{Wijesinghe et 
al.}{2012}]{Wijesinghe12} Wijesinghe D.~B., et al., 2012, MNRAS, 
423, 3679


\end{thebibliography}
\end{document}